\documentclass[11pt, a4paper]{article}
\pdfoutput=1
\usepackage{jcappub}

\usepackage{booktabs}
\usepackage{multirow}
\usepackage{amssymb}
\usepackage[export]{adjustbox}
\usepackage{floatrow}
\newfloatcommand{capbtabbox}{table}[][3.5in]
\newfloatcommand{capbfigbox}{figure}[][2.3in]
\usepackage{blindtext}
\usepackage{amsmath}
\usepackage{booktabs}
\usepackage{cancel}
\usepackage{aas_macros}

\newcommand{\ie}{i.e.~}

\def\lsim{\mathrel{\raise.3ex\hbox{$<$\kern-.75em\lower1ex\hbox{$\sim$}}}}
\def\gsim{\mathrel{\raise.3ex\hbox{$>$\kern-.75em\lower1ex\hbox{$\sim$}}}}

\usepackage{appendix}
\usepackage{array}
\newcolumntype{x}[1]{>{\centering\arraybackslash}p{#1}}
\usepackage{bm}
\usepackage{color}
\usepackage{graphicx}
\usepackage{cancel}
\usepackage{amsfonts}
\usepackage{amssymb}
\usepackage{amsmath}
\usepackage{calc}
\usepackage{dcolumn}
\usepackage{mathrsfs}
\usepackage{mathtools}
\usepackage{soul}

\usepackage{braket}
\usepackage[normalem]{ulem}

\newcommand{\Fig}[1]{Fig.~\ref{#1}}

\newcommand{\Lag}{\mathscr{L}}	

\newcommand{\beq}{\begin{equation}}
\newcommand{\eeq}{\end{equation}}

\definecolor{rossoCP3}{cmyk}{0,.88,.77,.40}
\definecolor{verdeCP3}{rgb}{0.09765625, 0.57421875, 0.1015625}
\definecolor{bluCP3}{rgb}{0, 0.23, 0.67}

\ifx\pdfoutput\undefined
\usepackage[dvips,bookmarks]{hyperref}    
\else
\usepackage{hyperref}    
\fi
\hypersetup{colorlinks, bookmarksopen, bookmarksnumbered,
citecolor=verdeCP3, linkcolor=bluCP3, pdfstartview=FitH, urlcolor=rossoCP3}

\begin{document}
\begin{flushright}
\large \tt IFIC/16-96 \\ FERMILAB-PUB-16-605-A 

\end{flushright}

\vskip 0.2in

\title{Updated Collider and Direct Detection Constraints on Dark Matter Models for the Galactic Center Gamma-Ray Excess}

\author[a,b]{Miguel Escudero,}\note{ORCID: http://orcid.org/0000-0002-4487-8742}
\emailAdd{miguel.escudero@ific.uv.es}
\author[b,c,d]{Dan Hooper}\note{ORCID: http://orcid.org/0000-0001-8837-4127}
\emailAdd{dhooper@fnal.gov}
\author[b,e]{and Samuel J.~Witte}\note{ORCID: http://orcid.org/0000-0003-4649-3085}
\emailAdd{switte@physics.ucla.edu}

\affiliation[a]{Instituto de F\'{\i}sica Corpuscular (IFIC)$,$ CSIC-Universitat de Val\`encia$,$ Apartado de Correos 22085$,$ E-46071 Valencia$,$ Spain}
\affiliation[b]{Fermi National Accelerator Laboratory, Center for Particle
Astrophysics, Batavia, IL 60510}
\affiliation[c]{University of Chicago, Department of Astronomy and Astrophysics, Chicago, IL 60637}
\affiliation[d]{University of Chicago, Kavli Institute for Cosmological Physics, Chicago, IL 60637}
\affiliation[e]{University of California, Los Angeles, Department of Physics and Astronomy, Los Angeles, CA 90095}

\abstract{Utilizing an exhaustive set of simplified models, we revisit dark matter scenarios potentially capable of generating the observed Galactic Center gamma-ray excess, updating constraints from the LUX and PandaX-II experiments, as well as from the LHC and other colliders. We identify a variety of pseudoscalar mediated models that remain consistent with all constraints. In contrast, dark matter candidates which annihilate through a spin-1 mediator are ruled out by direct detection constraints unless the mass of the mediator is near an annihilation resonance, or the mediator has a purely vector coupling to the dark matter and a purely axial coupling to Standard Model fermions. All scenarios in which the dark matter annihilates through $t$-channel processes are now ruled out by a combination of the constraints from LUX/PandaX-II and the LHC.}

\maketitle

\section{Introduction}

Over the past decade or so, a number of observations have been interpreted as possible signals of annihilating or decaying dark matter particles. Examples of such observations include the 511 keV emission from the Galactic Bulge~\cite{Boehm:2003bt}, an excess of synchrotron emission known as the WMAP Haze~\cite{Finkbeiner:2004us,Hooper:2007kb}, an excess of high energy positrons in the cosmic ray spectrum~\cite{Adriani:2008zr,Aguilar:2013qda}, a mono-energetic line of 130 GeV gamma rays from the Galactic Halo~\cite{Weniger:2012tx}, and a 3.5 keV X-ray line from Perseus and other galaxy clusters~\cite{Bulbul:2014sua,Boyarsky:2014jta}. Each of these anomalies, however, has either failed to be confirmed by subsequent measurements~\cite{Ackermann:2013uma,Aharonian:2016gzq}, or has been shown to be quite plausibly explained by astrophysical phenomena~\cite{Hooper:2008kg,Cholis:2013psa,Crocker:2016zzt,Su:2010qj}.

In comparison to these other anomalous signals, the gamma-ray excess observed from the Galactic Center by the Fermi Gamma-Ray Space Telescope stands out. This signal has been studied in detail over the past seven years~\cite{TheFermi-LAT:2015kwa,Daylan:2014rsa,Calore:2014xka,Hooper:2013rwa,Gordon:2013vta,Abazajian:2012pn,Hooper:2011ti,Hooper:2010mq,Goodenough:2009gk,Zhou:2014lva,Huang:2015rlu} and has been shown with high statistical significance to exhibit a spectrum, morphology and overall intensity that is compatible with that predicted from annihilating dark matter particles in the form of a $\sim$\,30-70 GeV thermal relic distributed with a profile similar to that favored by numerical simulations. Although astrophysical interpretations of this signal have been proposed (consisting of either a large population of millisecond pulsars~\cite{Cholis:2014lta,Lee:2015fea,Bartels:2015aea,Petrovic:2014xra,Hooper:2013nhl,Hooper:2015jlu,Hooper:2016rap,Brandt:2015ula}, or a series of recent leptonic cosmic-ray outbursts~\cite{Cholis:2015dea,Petrovic:2014uda,Carlson:2014cwa}), these explanations require either a significant degree of tuning in their parameters~\cite{Cholis:2015dea}, or pulsar populations which are very different from those observed in the environments of globular clusters or in the field of the Milky Way~\cite{Hooper:2016rap,Hooper:2015jlu,Cholis:2014lta}. In addition, some modest support for a dark matter interpretation of this signal has recently appeared in the form of excesses in the cosmic-ray antiproton spectrum~\cite{Cuoco:2016eej,Cui:2016ppb,Hooper:2014ysa}, in the gamma-ray emission from the dwarf spheroidal galaxies Reticulum II and Tucana III~\cite{Geringer-Sameth:2015lua,Drlica-Wagner:2015xua,Hooper:2015ula,Li:2015kag,Fermi-LAT:2016uux}, and from the observation of spatially extended gamma-ray emission from two dark matter subhalo candidates~\cite{Bertoni:2016hoh,Bertoni:2015mla,Xia:2016uog,Wang:2016xjx}. At this point in time, however, there is no clear resolution to the question of the origin of the Galactic Center excess.

\begin{table*}[htp]\label{table1}
\begin{center}
{\setlength{\extrarowheight}{2pt}
\begin{tabular}{|c|c|c|c|}
\hline \bf{ \emph{Dark Matter}} &  {\bf \emph{Mediator}} &  {\bf \emph{Interactions}} & {\bf Direct Detection}  \\ 
\hline \hline 
Dirac Fermion, $\chi$ & Spin-$0$  &   $\bar{\chi}\gamma^5\chi$,  $\bar{f}f$&  $\sigma_{\rm SI} \propto (q/2m_{\chi})^2$   \\ \hline
Majorana Fermion, $\chi$ & Spin-$0$ &  $\bar{\chi}\gamma^5\chi$,  $\bar{f}f$&  $\sigma_{\rm SI} \propto (q/2m_{\chi})^2$   \\ \hline
Dirac Fermion, $\chi$ & Spin-$0$  &  $\bar{\chi}\gamma^5\chi$,  $\bar{f}\gamma^5f$&  $\sigma_{\rm SD} \propto (q^2/4 m_n m_{\chi})^2$  \\ \hline
Majorana Fermion, $\chi$ & Spin-$0$ &  $\bar{\chi}\gamma^5\chi$,  $\bar{f}\gamma^5f$ & $\sigma_{\rm SD} \propto (q^2/4 m_n m_{\chi})^2$  \\ \hline 
Complex Scalar, $\phi$ & Spin-$0$  &    $\phi^{\dagger}\phi$,  $\bar{f}\gamma^5f$ &$\sigma_{\rm SD} \propto (q/2m_n)^2$  \\ \hline
Real Scalar, $\phi$        & Spin-$0$   &   $\phi^2 $,  $\bar{f}\gamma^5f$                     & $\sigma_{\rm SD} \propto (q/2m_n)^2$  \\ \hline
Complex Vector, $X$ & Spin-$0$  &  $X^{\dagger}_{\mu}X^{\mu}$,  $\bar{f}\gamma^5f$ & $\sigma_{\rm SD} \propto (q/2m_n)^2$  \\ \hline
Real Vector, $X$ & Spin-$0$     & $X_{\mu}X^{\mu}$,  $\bar{f}\gamma^5f     $ &       $\sigma_{\rm SD} \propto (q/2m_n)^2$       \\ 
\hline \hline
Dirac Fermion, $\chi$ & Spin-$1$ &    $\bar{\chi}\gamma^{\mu}\chi$, $\bar{b}\gamma_{\mu}b$ & $\sigma_{\rm SI} \sim$ loop (vector)  \\
\hline
Dirac Fermion, $\chi$ & Spin-$1$    & $\bar{\chi}\gamma^{\mu}\chi$,  $\bar{f}\gamma_{\mu}\gamma^5f$ &$\sigma_{\rm SD} \propto (q/2 m_n)^2$ or $(q/2 m_{\chi})^2$  \\ 
\hline
%
%
Dirac Fermion, $\chi$ & Spin-$1$ &    $\bar{\chi}\gamma^{\mu}\gamma^5\chi$,  $\bar{f}\gamma_{\mu}\gamma^5f$  & $\sigma_{\rm SD}\sim 1$ 
\\ \hline
Majorana Fermion, $\chi$ & Spin-$1$ &  $\bar{\chi}\gamma^{\mu}\gamma^5\chi$,  $\bar{f}\gamma_{\mu}\gamma^5f$ & $\sigma_{\rm SD}\sim$ 1  \\
\hline \hline
Dirac Fermion, $\chi$ & Spin-$0$ ($t$-ch.)    & $ \bar \chi  ( 1 \pm \gamma^5) b$  &   $\sigma_{\rm SI} \propto$ loop (vector)  \\ \hline
Dirac Fermion, $\chi$ & Spin-$1$ ($t$-ch.)  & $ \bar \chi \gamma^\mu ( 1 \pm \gamma^5) b$ & $\sigma_{\rm SI} \propto$ loop (vector) \\ \hline
Complex Vector, $X$ & Spin-$1/2$ ($t$-ch.) & $X_{\mu}^{\dagger} \gamma^{\mu}(1\pm \gamma^5) b$ & $\sigma_{\rm SI} \propto$  loop (vector) \\ \hline
Real Vector, $X$ & Spin-$1/2$ ($t$-ch.)& $X_{\mu} \gamma^{\mu}(1\pm \gamma^5) b$  & $\sigma_{\rm SI} \propto$ loop (vector)  \\ 
\hline \hline
\end{tabular}}
\end{center}
\caption{A summary of the simplified models identified in Ref.~\cite{Berlin:2014tja} as being potentially capable of generating the observed characteristics of the Galactic Center gamma-ray excess without violating collider or direct detection constraints (as of June 2014). For each model, we list the nature of the dark matter candidate and the mediator, as well as the form of the mediator's interactions. In the final column, we list whether the leading elastic scattering cross section with nuclei is spin-independent (SI) or spin-dependent (SD) and whether it is suppressed by powers of momentum, $q$, or by loops.}
\end{table*}%

Many groups have studied dark matter models that are capable of generating the observed features of the Galactic Center excess (for an incomplete list, see Refs.~\cite{Ipek:2014gua,Boehm:2014hva, Hooper:2014fda,Berlin:2014pya,Agrawal:2014una,Berlin:2014tja,Izaguirre:2014vva,Cheung:2014lqa,Cerdeno:2014cda,Alves:2014yha,Hooper:2012cw,Ko:2014gha,Boehm:2014bia,Abdullah:2014lla,Martin:2014sxa,Cline:2014dwa,Kim:2016csm,Karwin:2016tsw,Ghorbani:2014qpa}). In this study, we follow an approach similar to that taken in Ref.~\cite{Berlin:2014tja}, and consider an exhaustive list of simplified models that are capable of generating the observed gamma-ray excess while remaining consistent with all constraints from collider and direct detection experiments. Also following Ref.~\cite{Berlin:2014tja}, we choose to not consider hidden sector models in this study, in which the dark matter annihilates to unstable particles which reside in the hidden sector, without sizable couplings to the Standard Model (SM)~\cite{Abdullah:2014lla,Martin:2014sxa,Hooper:2012cw}. While such scenarios certainly remain viable, we consider them to be beyond the scope of this work. 

The models found in Ref.~\cite{Berlin:2014tja} to be compatible with existing constraints from direct detection and collider experiments are listed in Table~\ref{table1}, and can be divided into three categories. First, there are models in which the dark matter annihilates into SM quarks through the $s$-channel exchange of a spin-zero mediator with pseudoscalar couplings. These models allow for an unsuppressed ($s$-wave) low-velocity annihilation cross section while generating a cross section for elastic scattering with nuclei that is suppressed by either two or four powers of momentum, thus evading direct detection constraints. In the second class of models, the dark matter annihilates through the $s$-channel exchange of a vector boson. In this case, it was found that direct detection constraints could be evaded if the mediator couples axially with quarks or couples only to the third generation. Lastly, there are models in which the dark matter annihilates to $b$-quarks through the $t$-channel exchange of a colored and electrically charged mediator.

In this paper, we revisit this collection of dark matter models, applying updated constraints from the Large Hadron Collider (LHC) and other collider experiments, in addition to recent constraints from the direct detection experiments LUX~\cite{Akerib:2016vxi} and PandaX-II~\cite{Tan:2016zwf}. We find that many of the models previously considered within the context of the Galactic Center excess are now excluded by a combination of these constraints.


\section{Constraints}

In this section, we summarize the constraints that we apply in this study. In particular, we consider constraints from the LHC and other accelerators, as derived from searches for mono-X events with missing energy (where X denotes a jet, photon, or $Z$), di-jet resonances, di-lepton resonances, exotic Higgs decays, sbottom searches, and exotic upsilon decays~\cite{Khachatryan:2016zqb, Khachatryan:2015dcf, CMS:2016pod,CMS-PAS-SUS-16-016,ATLAS:2016cyf,Aaboud:2016tnv,Aaboud:2016uro,CMS-PAS-HIG-16-006,CMS-PAS-HIG-16-025,Barate:2003sz,Dolan:2014ska}. We also summarize the current status of direct searches for dark matter, including the recent constraints presented by the LUX~\cite{Akerib:2016vxi} and PandaX-II~\cite{Tan:2016zwf} Collaborations.

\subsection{LHC}
\label{LHC}

Searches at CMS and ATLAS provide some of the most stringent constraints on dark matter, as well as on the particles that mediate the interactions of dark matter. In this study, we consider the bounds from the LHC as applied to a wide range of simplified models, the most stringent of which arise from CMS searches for mono-jet+MET, di-jet resonances, di-lepton resonances, di-tau resonances, and sbottom searches. Although we also considered constraints from the ATLAS Collaboration, they were slightly less restrictive than those from CMS.

LHC limits are typically published in one of two ways: (1) assuming a particular model and choice of couplings, a limit is presented on the parameter space in the dark matter mass-mediator mass plane, or (2) a limit is presented on the product of the production cross section and the branching fraction for a particular process. In this study, we will present our results in terms of the mediator mass and the product of the dark matter-mediator and SM-mediator couplings. Thus applying limits from the LHC generally requires translating these bounds into the parameter space under consideration. To calculate the relevant production cross sections and branching ratios, models are built using FeynRules~\cite{Alloul:2013bka} and subsequently imported into MadGraph5\_aMC@NLO~\cite{Alwall:2014hca, Hirschi:2015iia}. When necessary, we implement PYTHIA 8~\cite{Sjostrand:2007gs} to hadronize the final state particles and DELPHES~\cite{deFavereau:2013fsa} to simulate the detector response. As appropriate, we apply the published cuts on MET, final state momentum, and final state rapidity in our calculations. Throughout this study, we calculate and present all LHC constraints at the 95\% confidence level. 

In scenarios with heavy mediators, it is not uncommon for the width of the mediator to be unacceptably large (\ie as large or larger than its mass). Such widths are clearly not physical and may indicate the presence of additional particles or interactions~\cite{Duerr:2016tmh,Englert:2016joy,Boveia:2016mrp,Kahlhoefer:2015bea,Abdallah:2015ter}. Imposing unitarity and gauge invariance often restricts the mass of such additional particles to be of the same order of magnitude as the other dark sector particles, making it difficult to define the properties of these new particles such that they are beyond the reach of the LHC. Although the construction of more complicated dark sectors is beyond the scope of the work, we emphasize that it is likely that constraints derived on such models would be more restrictive than those derived here. Throughout this study, in order to maintain the validity of the theory in this region of parameter space, we apply LHC constraints assuming $\Gamma/m = 0.1$ whenever the width of the mediator would otherwise exceed this value.

\subsection{LEP-II}

Constraints from LEP-II on Higgs bosons in the mass range between $10$ GeV and $100$ GeV are extremely constraining for a wide range of beyond the SM physics scenarios. In this study we consider such limits as derived from searches for a light Higgs decaying to $b\bar{b}$~\cite{Barate:2003sz}. Although powerful, these constraints are rather model dependent, and generally rely on the scalar mediator's coupling to the SM gauge bosons. LEP-II constraints are presented at the 95\% confidence level throughout this work, and assume a coupling to the Z-boson identical to that of the SM Higgs.

\subsection{BaBar}

We also consider in this study constraints derived from BaBar on upsilon decays to light scalar or pseudoscalar particles, in particular focusing on channels where the mediator subsequently decays to hadrons, muons, taus or charm quarks~\cite{Lees:2011wb,Lees:2012te,Lees:2012iw,Lees:2015jwa}. We consider relativistic and QCD corrections for the decay of a vector meson as described in Ref.~\cite{Dolan:2014ska}. We note that the $\mu^+ \mu^-$ channel provides the strongest constraints, but the precise values of the branching ratios of such light scalars are not well known (see e.g.~Refs.~\cite{Dolan:2014ska,Clarke:2013aya}). Here, we conservatively assume a 100\% branching ratio to hadrons in the mass range of $1 \, {\rm GeV} \lesssim m_{A} \lesssim 2 m_\tau$. This is conservative in the sense that introducing a small branching ratio to muons strengthens the resulting bound. For $ 2 m_\tau \lesssim m_{A} < 9 \, {\rm GeV}$, we use the branching ratios as recently computed in Ref.~\cite{Haisch:2016hzu} which incorporate QCD corrections. We find similar constraints as those previously obtained in the recent analysis of Refs.~\cite{Dolan:2014ska} and \cite{Clarke:2013aya} for pseudoscalar and scalar mediators, respectively. All BaBar constraints are presented at the 90\% confidence level in this study.

\subsection{Direct Detection}

The constraints utilized in this study on the dark matter's elastic scattering cross section with nuclei have been derived from the latest results of the LUX Collaboration~\cite{Akerib:2016vxi}, which are only slightly more stringent than those recently presented by the PANDA-X experiment~\cite{Tan:2016zwf}. 

For all tree-level cross sections, we use the expressions as presented in Appendices B and C of Ref.~\cite{Berlin:2014tja}. One-loop cross sections for the scalar mediated $t$-channel interaction and the $s$-channel vector mediated loop-suppressed interaction are provided in Refs.~\cite{Agrawal:2014una} and~\cite{Kopp:2009et}, respectively. The remaining $t$-channel models, which are also loop suppressed, suffer from the problem that they are not generically gauge invariant. Consequently, scattering cross sections for these models are calculated by introducing a factor that suppresses the cross section by the same factor that would appear if the interaction were mediated by a massive photon, \ie $\left(g^2 \log(m_b^2/m_{\rm med}^2) /(64\pi^2m_{\rm med}^2)\right)^2$.

For each model, we calculate the expected number of events in a xenon target following the procedure outlined in Ref.~\cite{Savage:2008er}, adopting a standard Maxwellian velocity distribution ($v_0 = 220$ km/s, $v_{\rm esc} = 544$ km/s, $\bar{v}_{\rm Earth}=245$ km/s), a local density of $0.3 \, {\rm GeV/cm^3}$ and an exposure of $3.35\times 10^{4}$ kg-day. Form factors and nuclear responses are calculated following the procedures outlined in Refs.~\cite{Cheng:2012qr,Fitzpatrick:2012ix}. We take the efficiency for nuclear recoils as a function of energy from Fig.~2 of Ref.~\cite{Akerib:2016vxi}, and derive bounds at the 90\% confidence level, assuming $4.2$ expected background events and applying Poisson statistics.


 

\section{Pseudoscalar Mediated Dark Matter}\label{spin0med}

In this section, we will consider models in which the dark matter annihilates through the $s$-channel exchange of a spin-0 mediator, $A$. We begin by considering a fermionic dark matter candidate, $\chi$, with interactions as described by the following Lagrangian:    
\begin{equation}
\Lag \supset \left[ a\bar{\chi} \lambda_{\chi p}i \gamma^{5} \chi + \sum_{f} y_{f}\bar{f}(\lambda_{f s} + \lambda_{f p}i \gamma^{5})f \right] A \, ,
\end{equation}
where $a = 1 (1/2)$ for a Dirac (Majorana) dark matter candidate. Although we describe the interactions of the SM fermions in terms of their yukawas, $y_f \equiv \sqrt{2} m_f / v$, the quantities $\lambda_{fs}$ and $\lambda_{fp}$ allow for arbitrary values of each coupling. Here, $v$ is the SM Higgs vacuum expectation value, \ie $v \simeq 246$ GeV. Assuming that $\lambda_{bs}$ or $\lambda_{bp}$ is not much smaller than that of the other SM fermions, dark matter will annihilate largely to $b\bar{b}$ in this model. For this dominant annihilation channel, a dark matter mass of approximately 50 GeV is required to generate the observed spectral shape of the Galactic Center excess~\cite{Calore:2014xka,Calore:2014nla}, and we adopt this value throughout this section.

\begin{figure*}
\center
\includegraphics[width=0.49\textwidth]{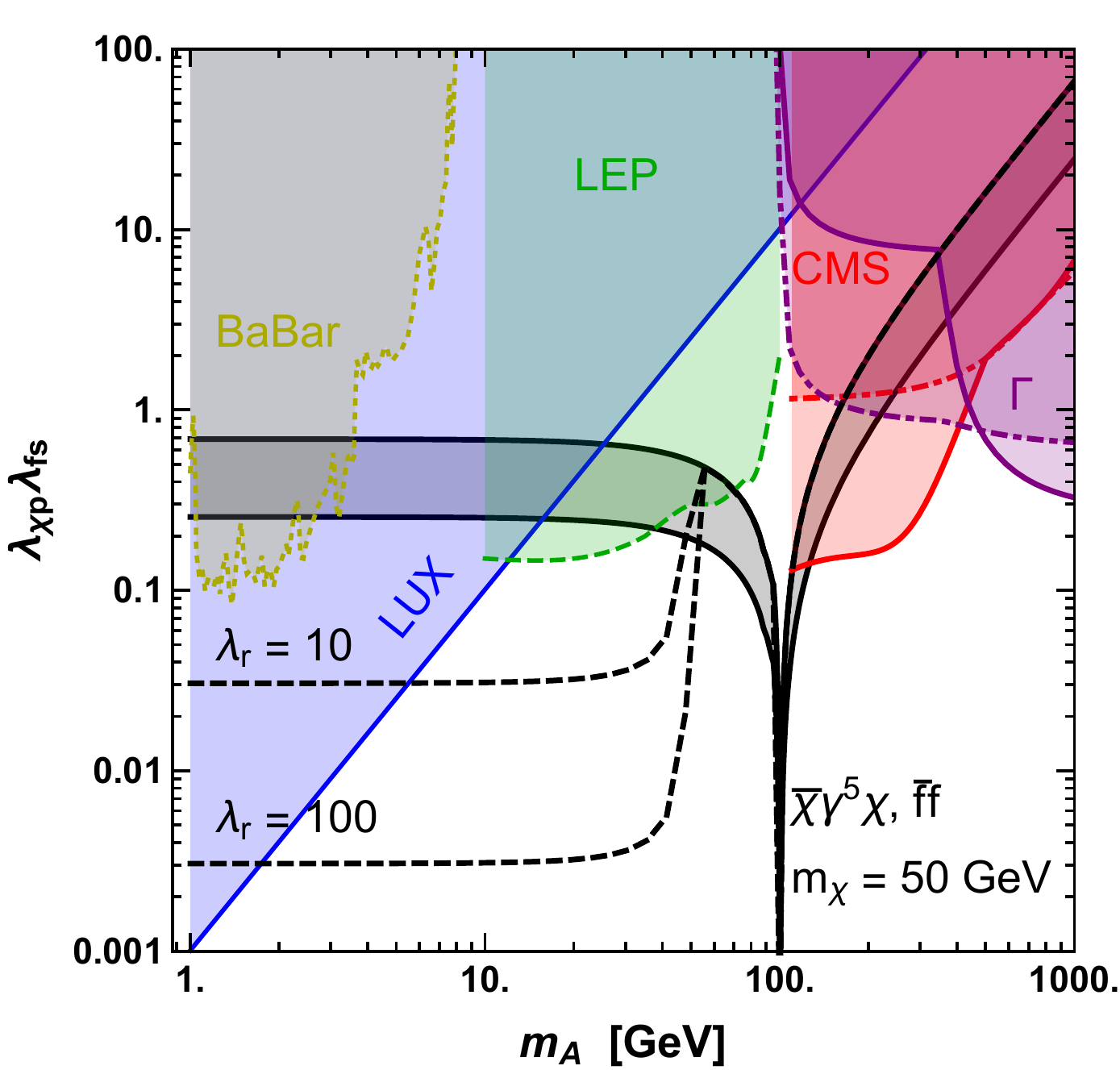}
\includegraphics[width=0.49\textwidth]{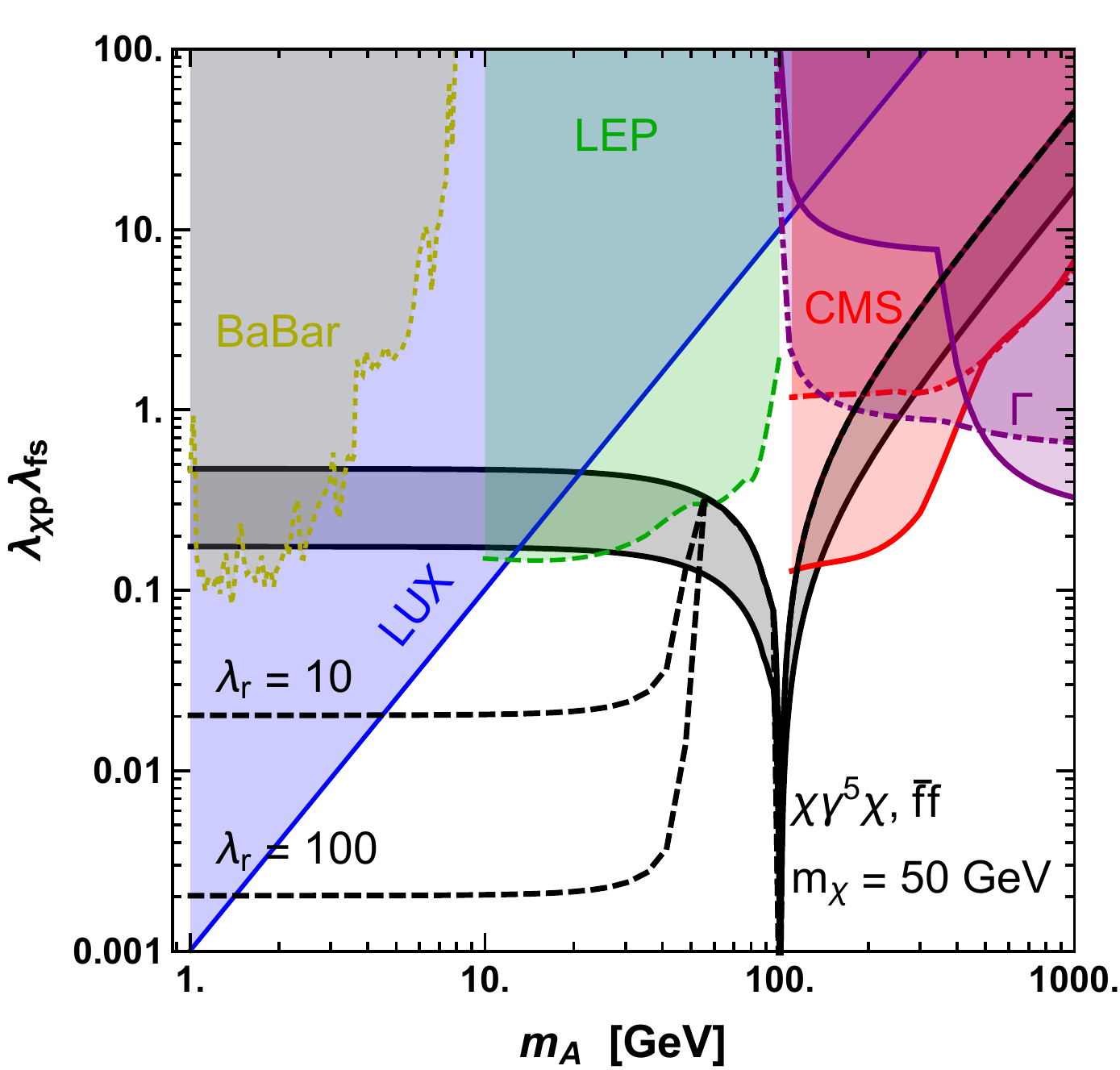}
\caption{\label{fig:spin0med_ps_s}Constraints on a $50$ GeV Dirac (left) and Majorana (right) dark matter candidate which annihilates through a spin-$0$ mediator with a pseudoscalar coupling to the dark matter and a (universal) scalar coupling to SM fermions. The black dashed (solid) lines include (neglect) annihilations to mediator pairs for several values of $\lambda_r \equiv \lambda_{\chi}/\lambda_{b}$. The upper boundary of the shaded black region is where the correct thermal relic abundance is obtained, whereas along the lower boundary the low-velocity annihilation cross section is at its minimum value required to potentially generate the observed gamma-ray excess. The constraints from CMS assume $\lambda_r = 1/3$ (solid) and $\lambda_r = 3$ (dash-dot), and are compared with the bounds enforcing $\Gamma_A / m_A = 0.1$ (purple) for the same coupling ratios. LEP and BaBar constraints are presented for $\lambda_r = 10$ and 1, respectively.}
\end{figure*}

\begin{figure*}
\center
\includegraphics[width=0.49\textwidth]{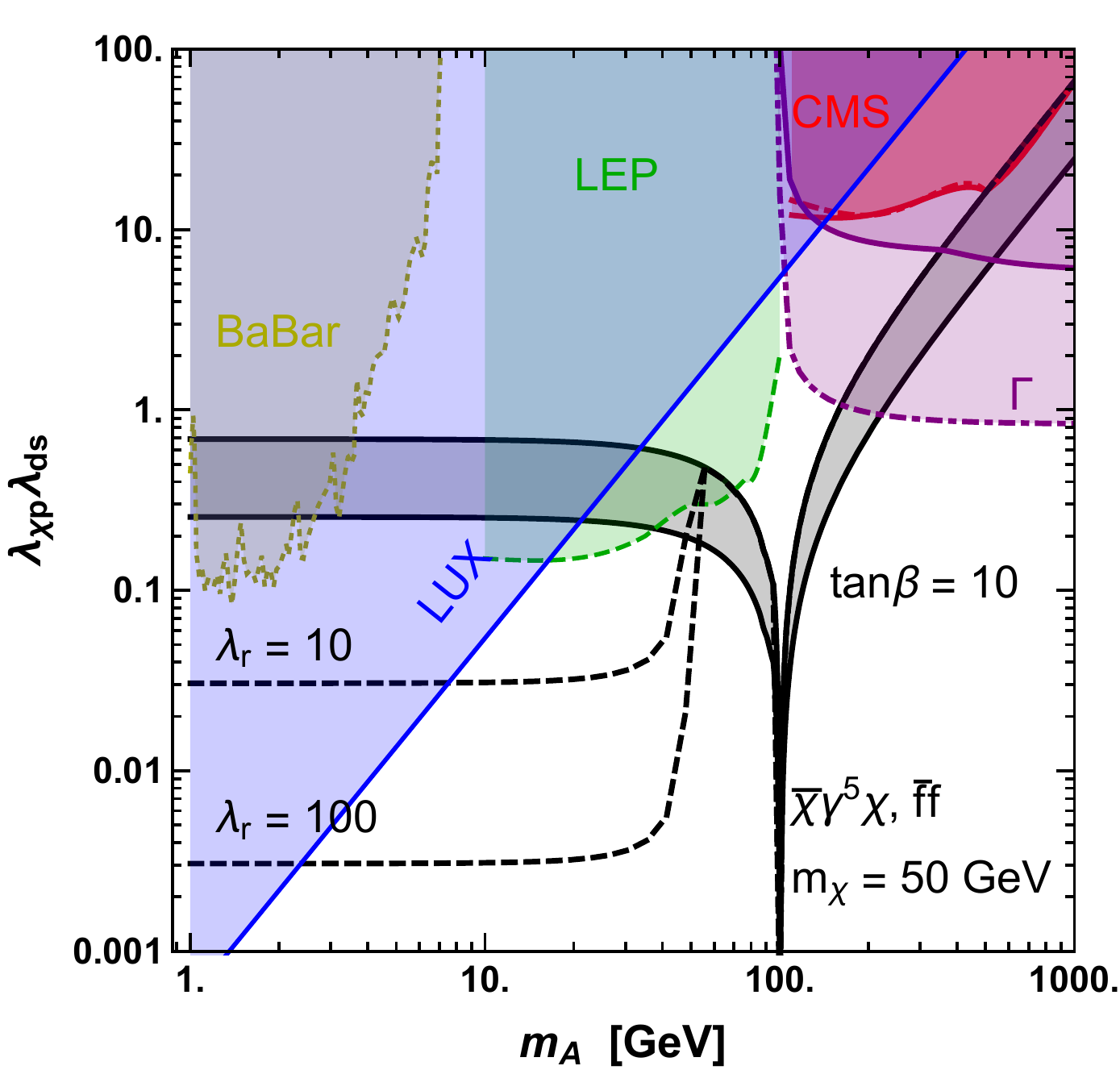}
\includegraphics[width=0.49\textwidth]{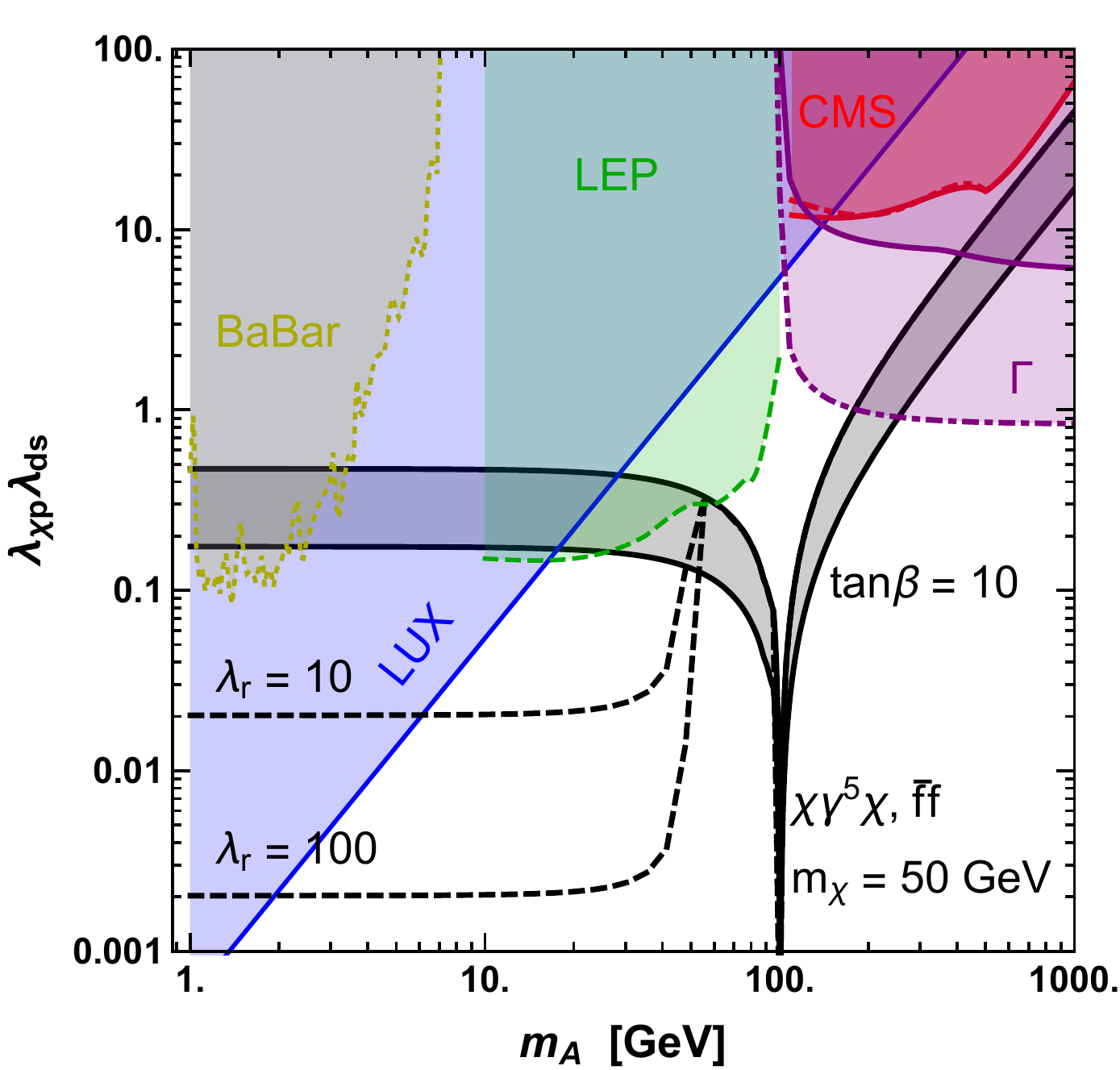}
\caption{\label{fig:spin0med_ps_s_tbeta} As in \Fig{fig:spin0med_ps_s} but for $\tan \beta = 10$, where $\tan \beta$ is defined as the ratio of the mediator's couplings to down-type and up-type fermions.}
\end{figure*}

In the left (right) frame of Fig.~\ref{fig:spin0med_ps_s}, we plot the constraints on the parameter space of a simplified model with dark matter in the form of a Dirac (Majorana) fermion and a mediator with a pseudoscalar coupling to the dark matter ($\bar{\chi} \gamma^5 \chi$) and scalar couplings to SM fermions ($\bar{f}f$), assuming a common scalar coupling to all SM fermions (as motivated by minimal flavor violation), $\lambda_{fs}$\footnote{Note that the product of couplings in these models be quite large, occasionally appearing to violate perturbativity. This need not be the case, however, as we have not included the yukawa contribution to the SM coupling, which may significantly suppress this product.}. In each frame, the upper boundary of the shaded black region represents the the value of the product of the couplings that is required to generate an acceptable thermal relic abundance, assuming standard cosmology. The lower boundary of this region corresponds to a more relaxed assumption, requiring only that the low-velocity annihilation cross section is large enough to potentially generate the observed gamma-ray excess, $\langle \sigma_{_{\chi \chi}} v \rangle > 3 \times 10^{-27}$ cm$^3/$s (or twice this value in the case of a Dirac particle). If $m_A < m_\chi$, dark matter can annihilate directly to mediator pairs via t-channel $\chi$ exchange. In these figures, we plot as dashed black lines the parameter space which generates the observed thermal relic abundance for several values of $\lambda_r \equiv \lambda_{\chi}/ \lambda_{b}$. One should keep in mind that if the dark matter annihilates significantly to mediator pairs in the low-velocity limit, a higher value for the dark matter mass is generally required in order for the resulting gamma-ray spectrum to be consistent with the observed features of the Galactic Center excess~\cite{Berlin:2014pya,Abdullah:2014lla,Martin:2014sxa}. We compare these curves to the constraints derived from LUX (blue), CMS/LHC (red), LEP (green), and BaBar (yellow).

In the case of CMS, the most stringent constraint in this class of models derives from searches for events with a single jet and missing transverse energy (MET).  As the sensitivity of collider searches depends not only on the product of the couplings, but also on their ratio, we present constraints for multiple values of $\lambda_r$. In Fig.~\ref{fig:spin0med_ps_s}, the solid (dot-dashed) lines correspond to CMS constraints for $\lambda_r= 1/3$ $(3)$, while LEP and BaBar constraints are derived assuming $\lambda_r = 10$ and $\lambda_r = 1$, respectively. The regions bounded by a purple solid (dot-dashed) line represent those in which the calculated width of the mediator exceeds one tenth of its mass, for $\lambda_r= 1/3$ $(3)$. As described in Sec.~\ref{LHC}, we set $\Gamma_A = 0.1 \,m_A$ throughout this region of parameter space, and take this to be indicative of additional particles and/or interactions that are not described by our simplified model.

The constraints from LEP rely on an effective coupling of the SM $Z$ to $ZA$, and are thus highly model dependent. While this constraint does apply, for example, to the case in which the couplings of the $A$ to SM fermions are the result of mixing with the SM Higgs, there are many other scenarios in which a spin-0 mediator can couple to the SM fermions while having a suppressed coupling to the $Z$.

%


Several of the constraints shown in Fig.~\ref{fig:spin0med_ps_s} depend on the ratios of the various couplings of the mediator. In particular, since the LHC constraints are dominated by diagrams in which a scalar mediator is produced through a top quark loop, such constraints may be much weaker if the top quark coupling is suppressed. To illustrate this, we plot in \Fig{fig:spin0med_ps_s_tbeta} the derived constraints assuming $\tan \beta  =10$, where $\tan \beta$ is defined as the ratio of the mediator's couplings to down-type and up-type fermions, $\tan \beta  \equiv \lambda_{d} / \lambda_{u}$. While bounds from LEP, LUX and BaBar are not significantly affected by the value of $\tan \beta$, mono-jet+MET bounds can be noticeably reduced, in particular in the case of $\lambda_r \ll 1$. Increasing $\tan \beta$ also reduces the width of the mediator for $m_A > 2 m_t$, potentially opening up additional parameter space.




\begin{figure*}
\center
\includegraphics[width=0.49\textwidth]{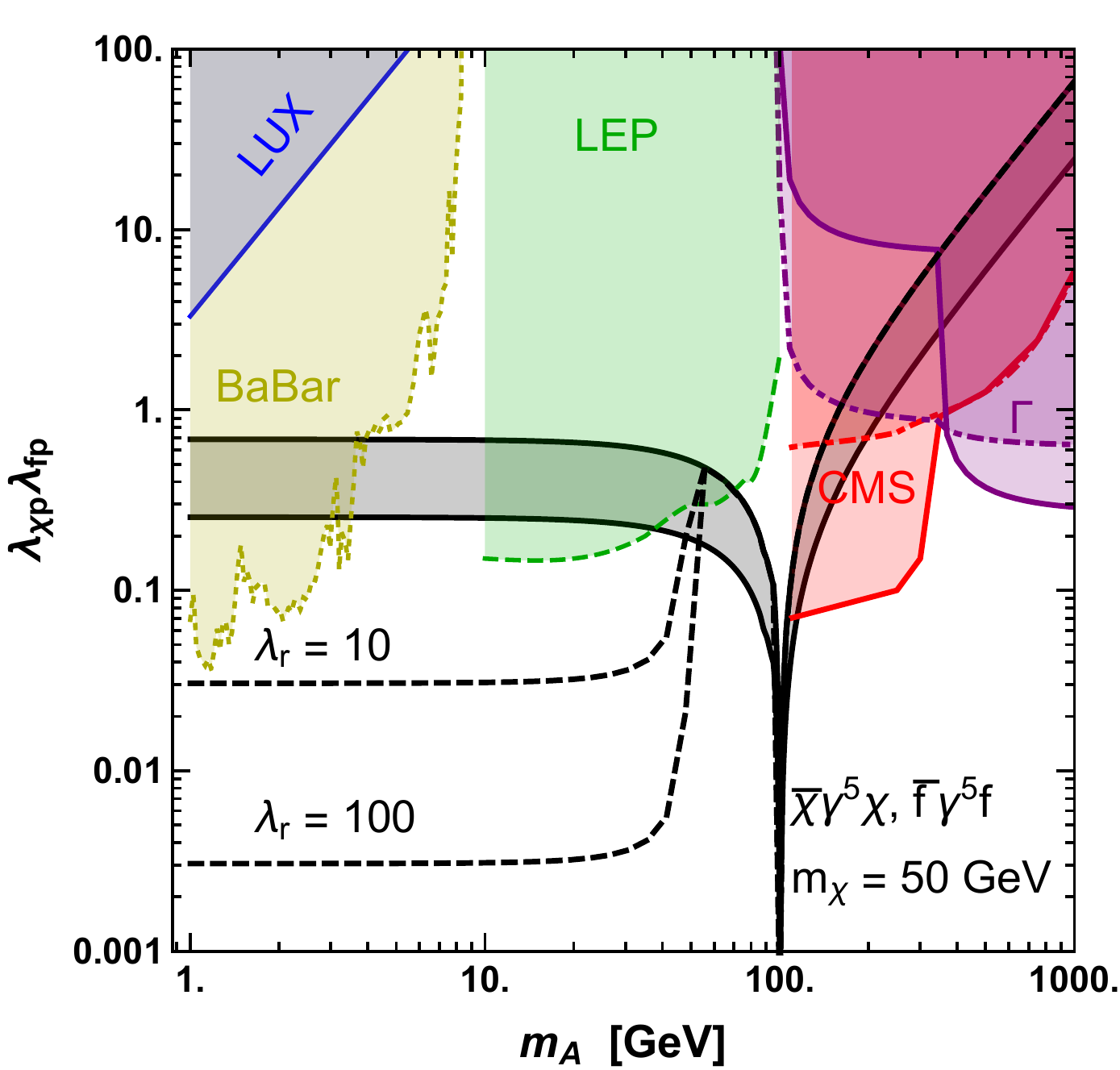}
\includegraphics[width=0.49\textwidth]{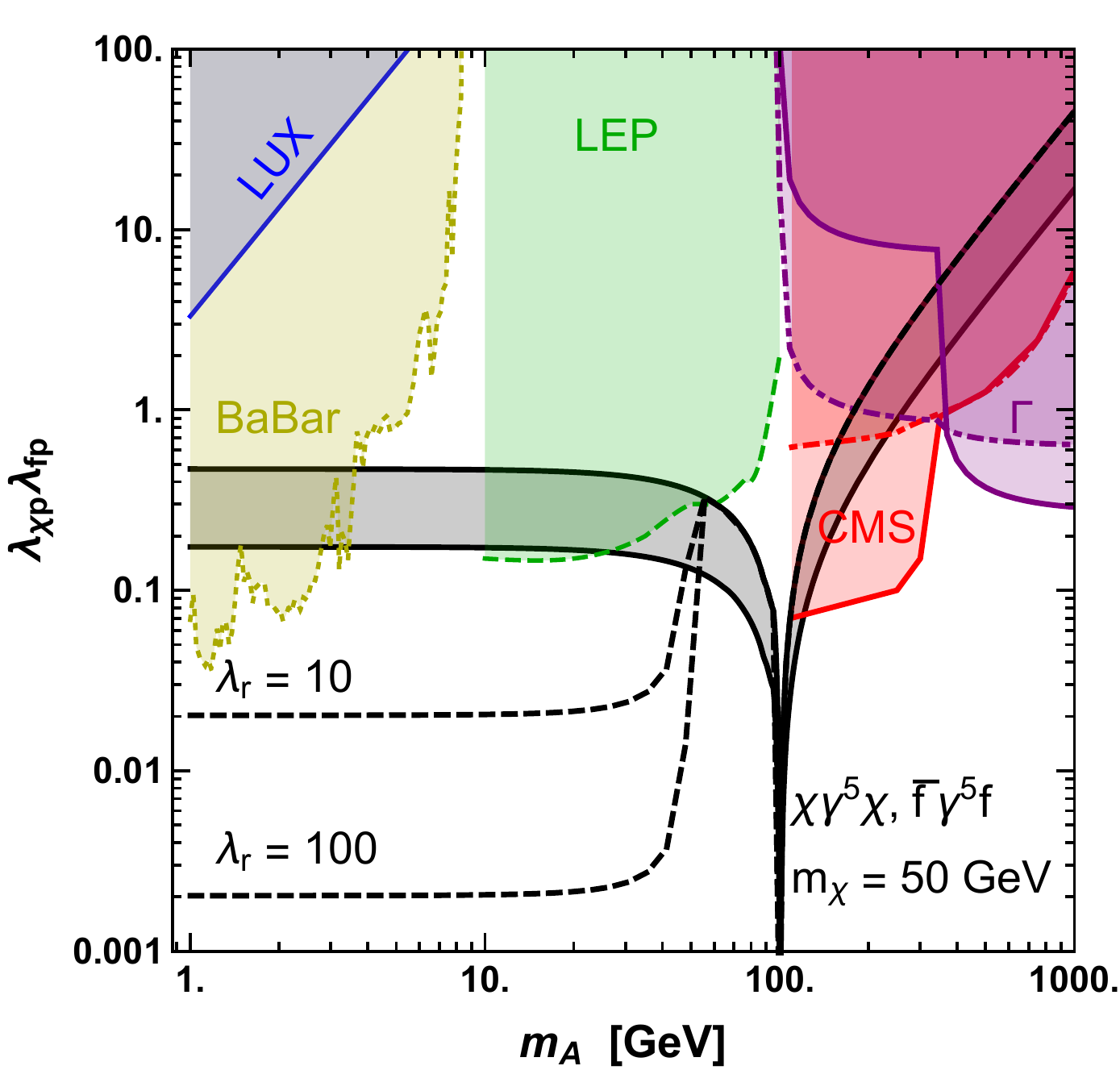} \\
\includegraphics[width=0.49\textwidth]{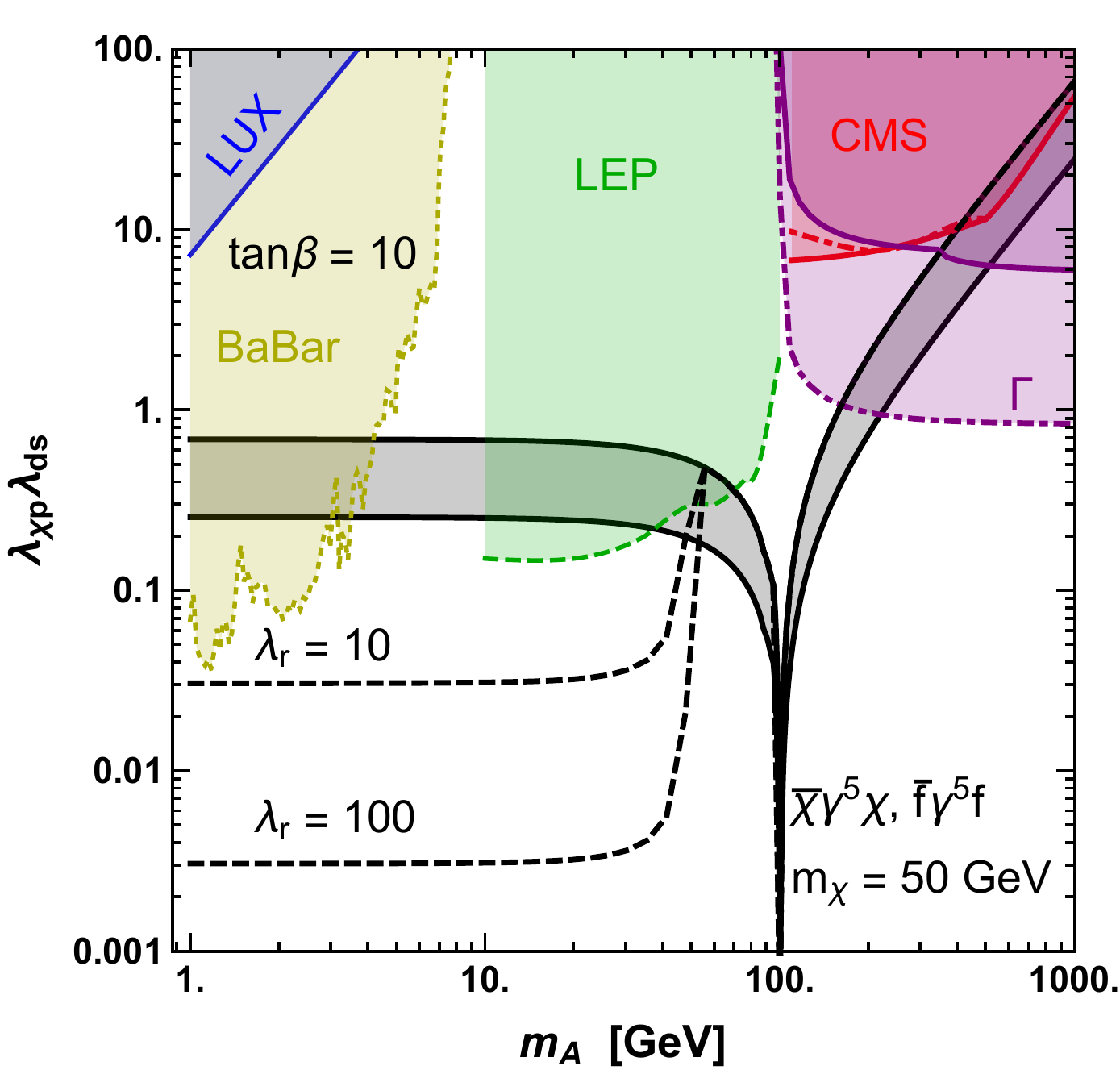}
\includegraphics[width=0.49\textwidth]{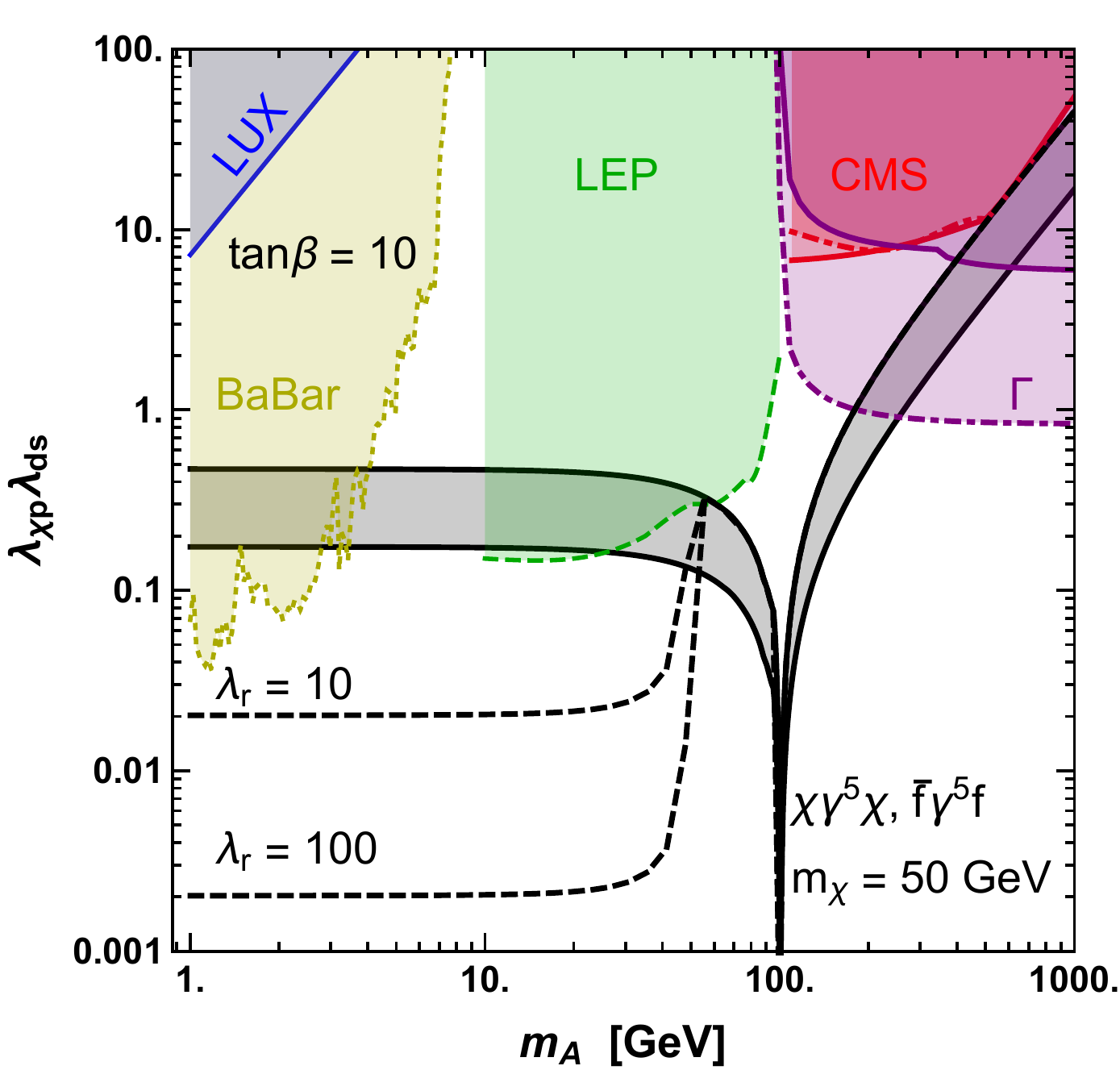}
\caption{\label{fig:spin0med_ps_ps}As in \Fig{fig:spin0med_ps_s} but for a mediator with purely pseudoscalar couplings. The upper (lower) frames correspond to $\tan \beta = 1$ (10).}
\end{figure*}

We repeat this exercise in Fig.~\ref{fig:spin0med_ps_ps} for the case of a mediator with pseudoscalar couplings to both the dark matter and to SM fermions. In this case, the dark matter's elastic scattering cross section with nuclei is both spin-dependent and heavily momentum suppressed ($\sigma_{SD} \propto q^4$), making direct detection experiments largely insensitive to these models. The bounds derived from colliders, however, are relatively insensitive to whether SM fermions couple via a scalar or pseudoscalar interaction. We emphasize that, as in the previous case, a large portion of parameter space remains viable for this model, especially should the top-mediator coupling be suppressed.




Next, we consider dark matter in the form of a scalar $\phi$, with a Lagrangian given by: 
\begin{equation}
\Lag \supset \left[ a \mu_\phi |\phi|^2 + \sum_{f} y_{f}\bar{f} \lambda_{f p}i \gamma^{5} f \right] A \, ,
\end{equation}
where $a = 1 (1/2)$ for a complex (real) dark matter particle.

\begin{figure*}
\center
\includegraphics[width=0.49\textwidth]{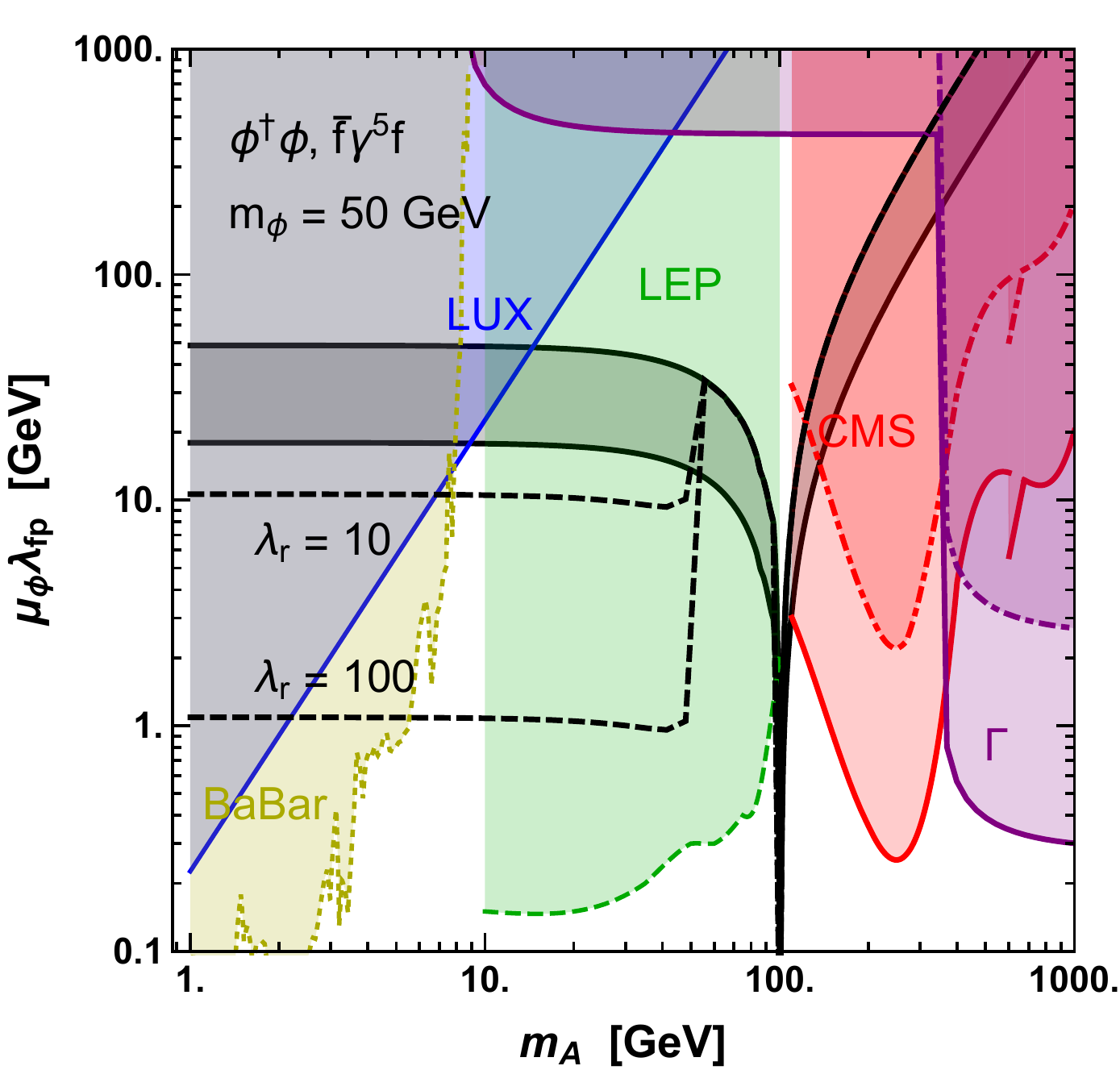}
\includegraphics[width=0.49\textwidth]{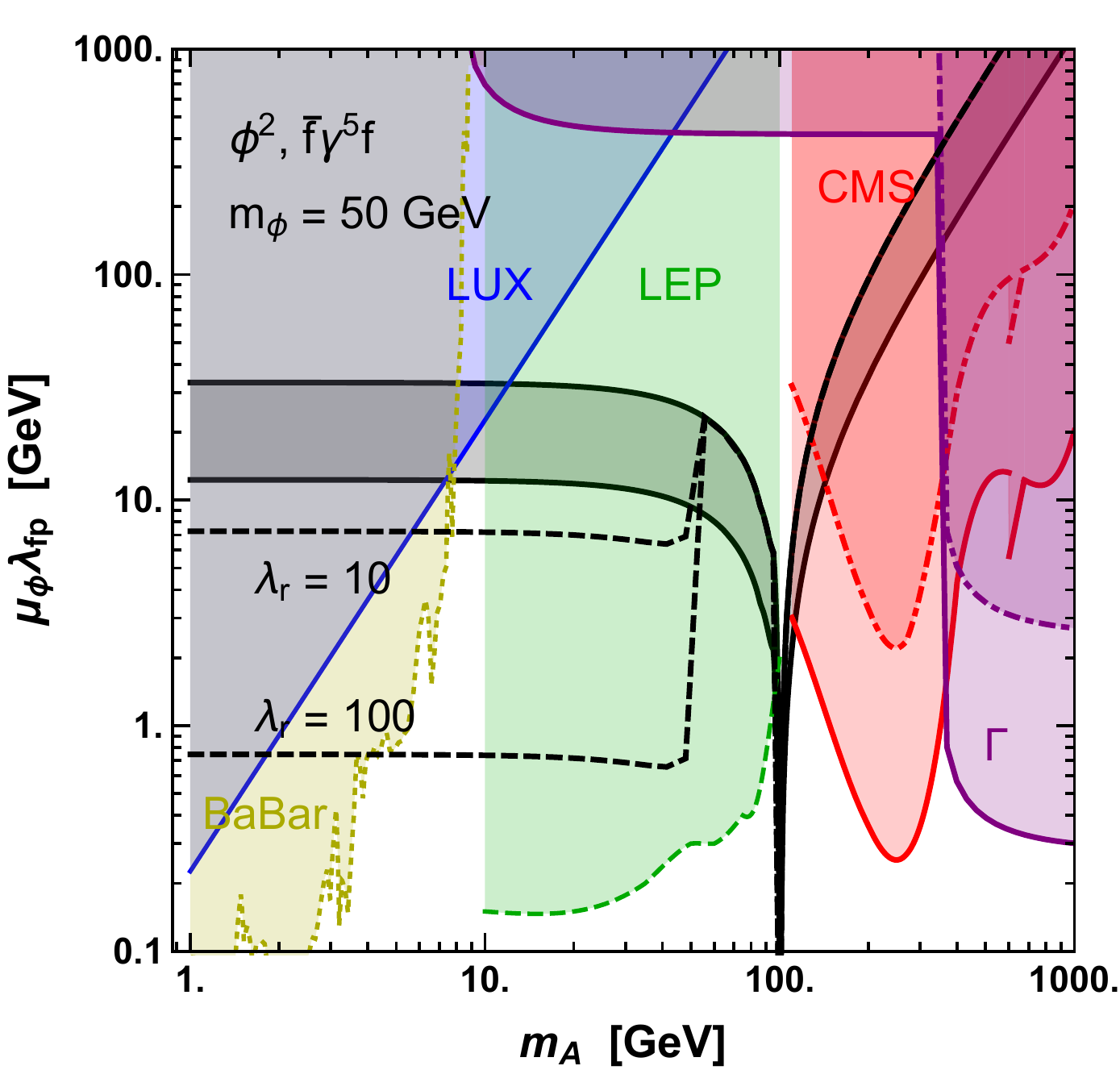} \\
\includegraphics[width=0.49\textwidth]{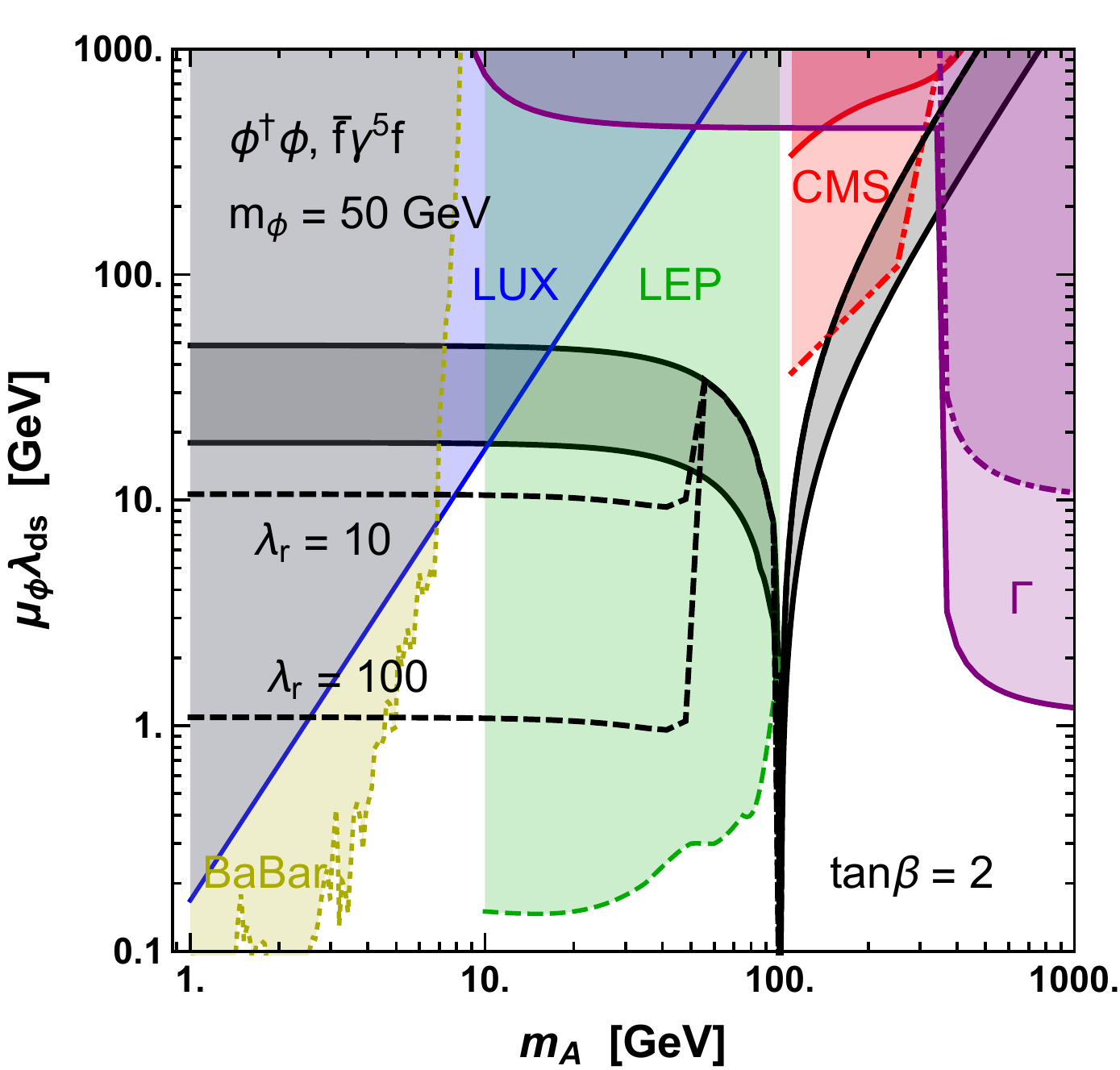}
\includegraphics[width=0.49\textwidth]{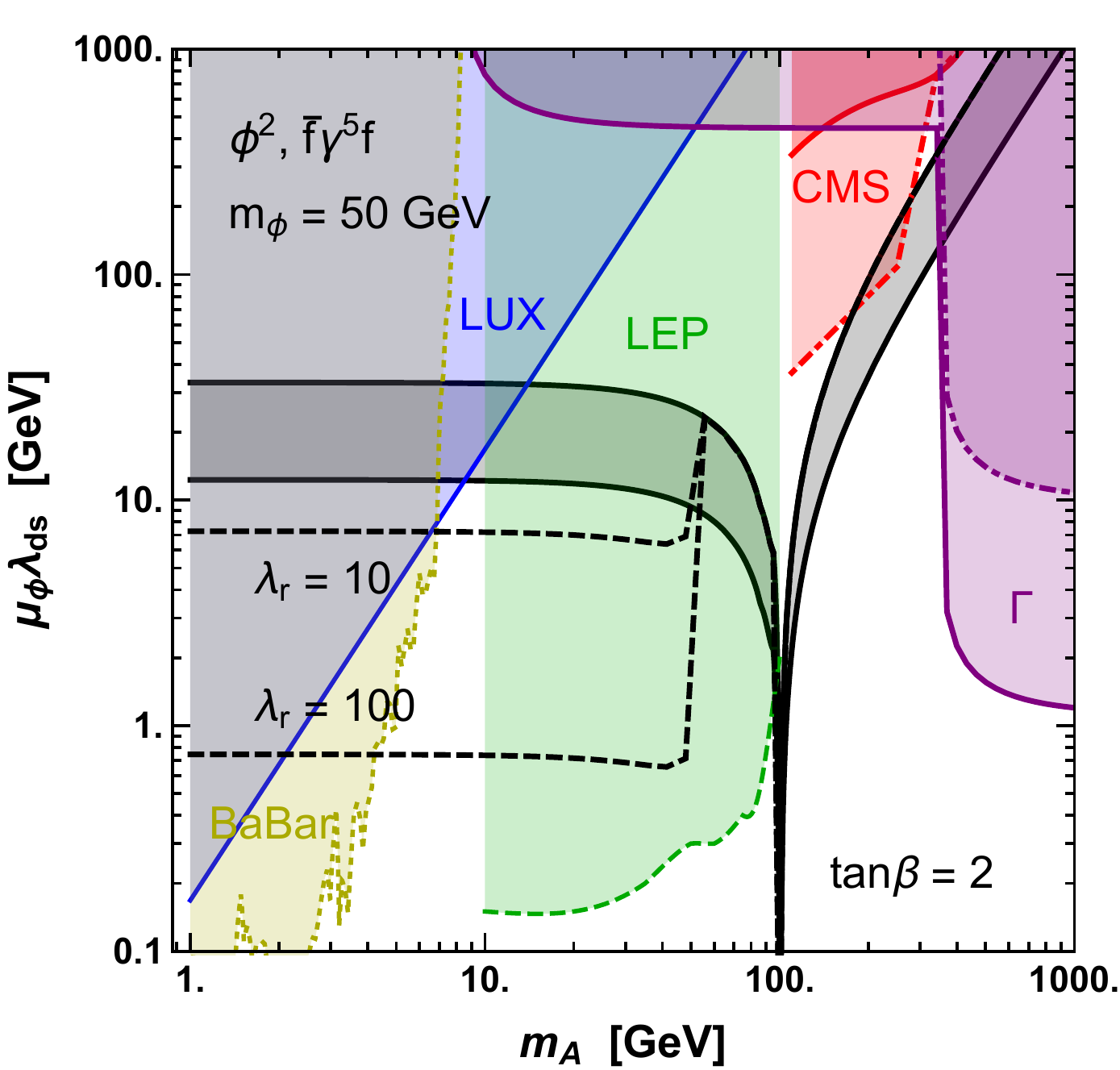}
\caption{\label{fig:spin0med_scalardm} As in the previous figures, but for a $50$ GeV complex (left) or real (right) scalar dark matter candidate, which annihilates through a spin-$0$ mediator with a pseudoscalar coupling to SM fermions. In the upper frames, we take the mediator's couplings to be equal for all SM fermions, whereas in the lower frames the mediator does not couple to leptons and $\tan \beta = 2$.}
\end{figure*}

The phenomenology of this model is summarized in \Fig{fig:spin0med_scalardm}, for the cases of a complex (left frame) or real (right frame) scalar. LHC signatures for this model are rather different from in the case of fermionic dark matter as the decay of the spin-$0$ mediator to dark matter is heavily suppressed. Instead, the dominant constraint from the LHC results from searches for a Higgs-like particle decaying to $\tau^+\tau^-$. At very large mediators masses, however, ($m_A \gtrsim 600$ GeV), the branching ratio to $\tau^+\tau^-$ is reduced and di-jet resonances become slightly more constraining (this accounts for the dip-like feature appearing in the CMS bound). As in the previous scenarios, LEP bounds on scalar decays to $b\bar{b}$ are very constraining in the region $10 \, \rm{GeV} < m_A < 100$ GeV, but only apply in models in which the mediator couples either directly or indirectly to the $Z$. 


In the lower frames of \Fig{fig:spin0med_scalardm} , we show how these bounds change if the mediator does not couple to leptons and has an asymmetric coupling to up-like and down-like quarks with $\tan \beta = 2$. This choice can open a window of parameter space for $100 \, {\rm GeV} \lesssim m_A \lesssim 2 m_{t}$, depending on the precise values of $\tan \beta$ and $\lambda_r$. 




\begin{figure*}
\center
\includegraphics[width=0.49\textwidth]{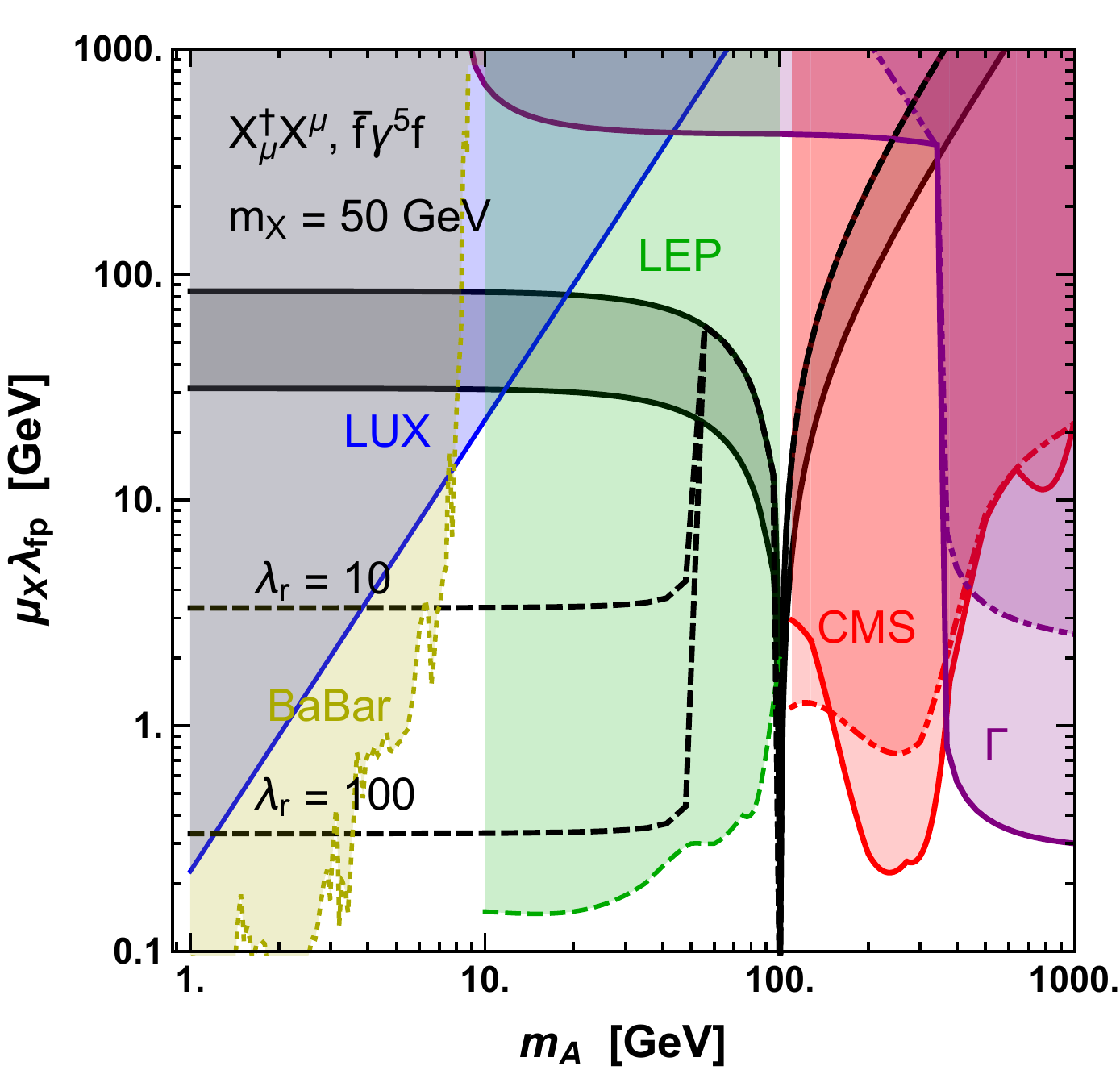} 
\includegraphics[width=0.49\textwidth]{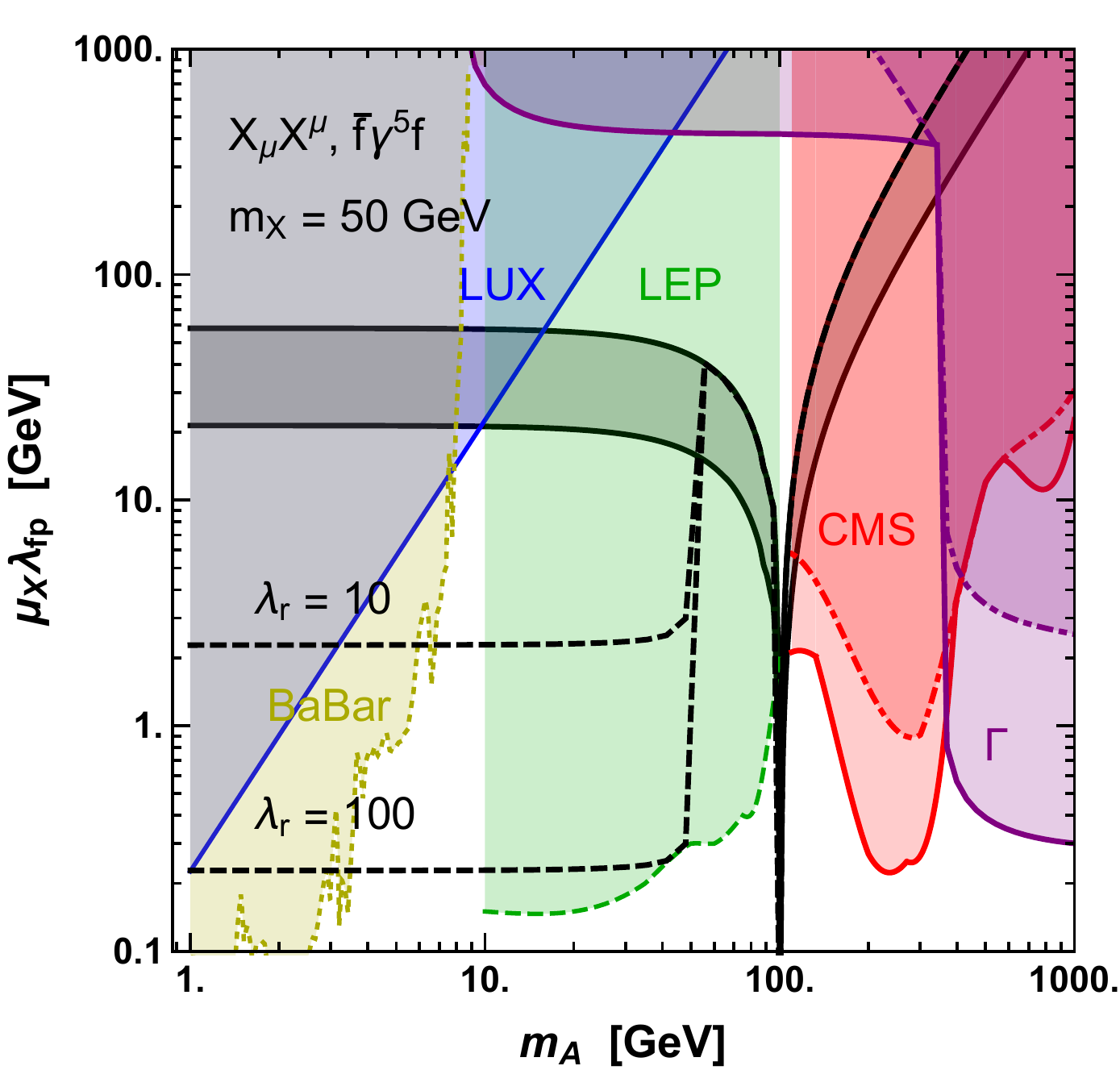}
\\
\includegraphics[width=0.49\textwidth]{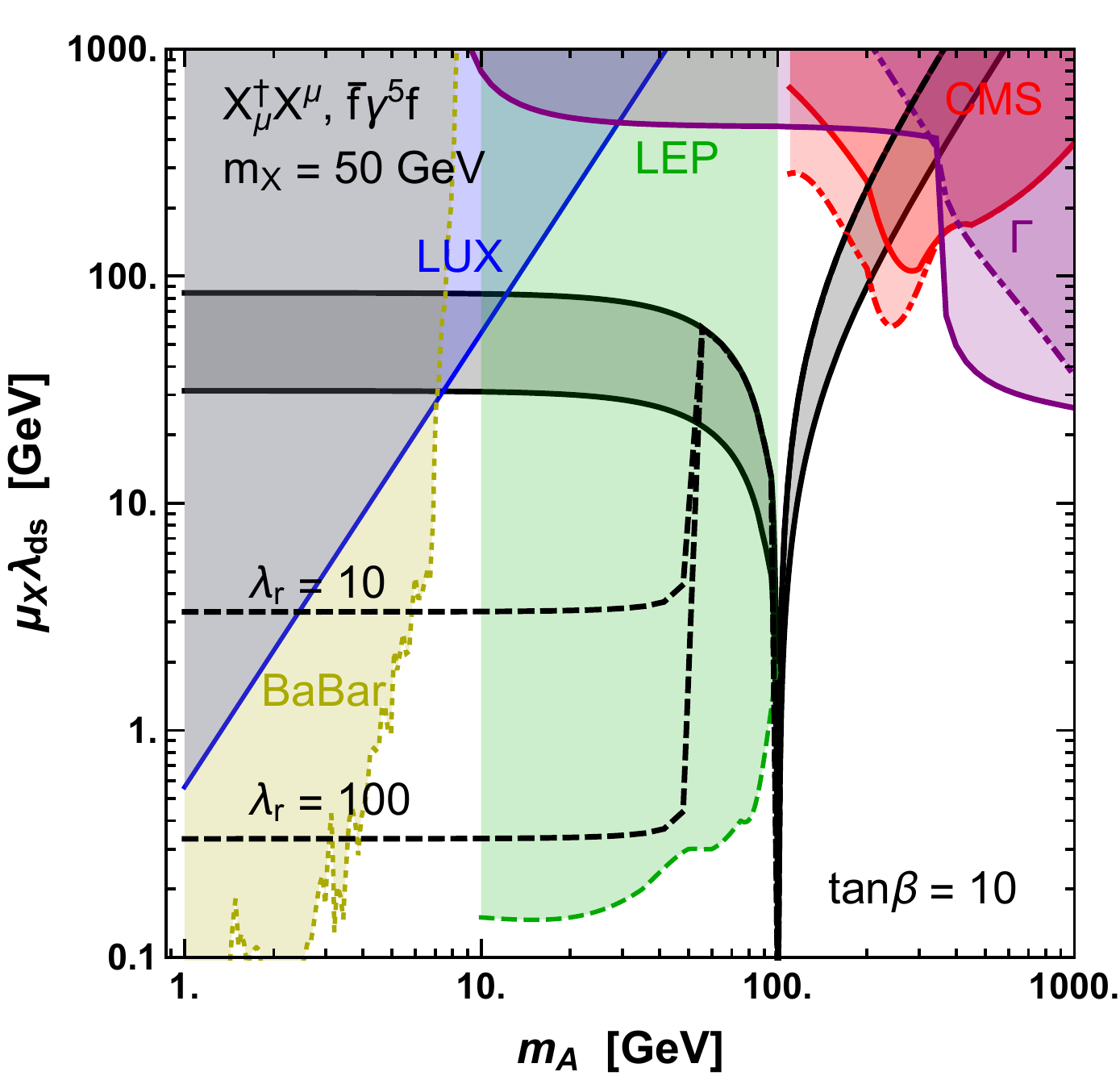}
\includegraphics[width=0.49\textwidth]{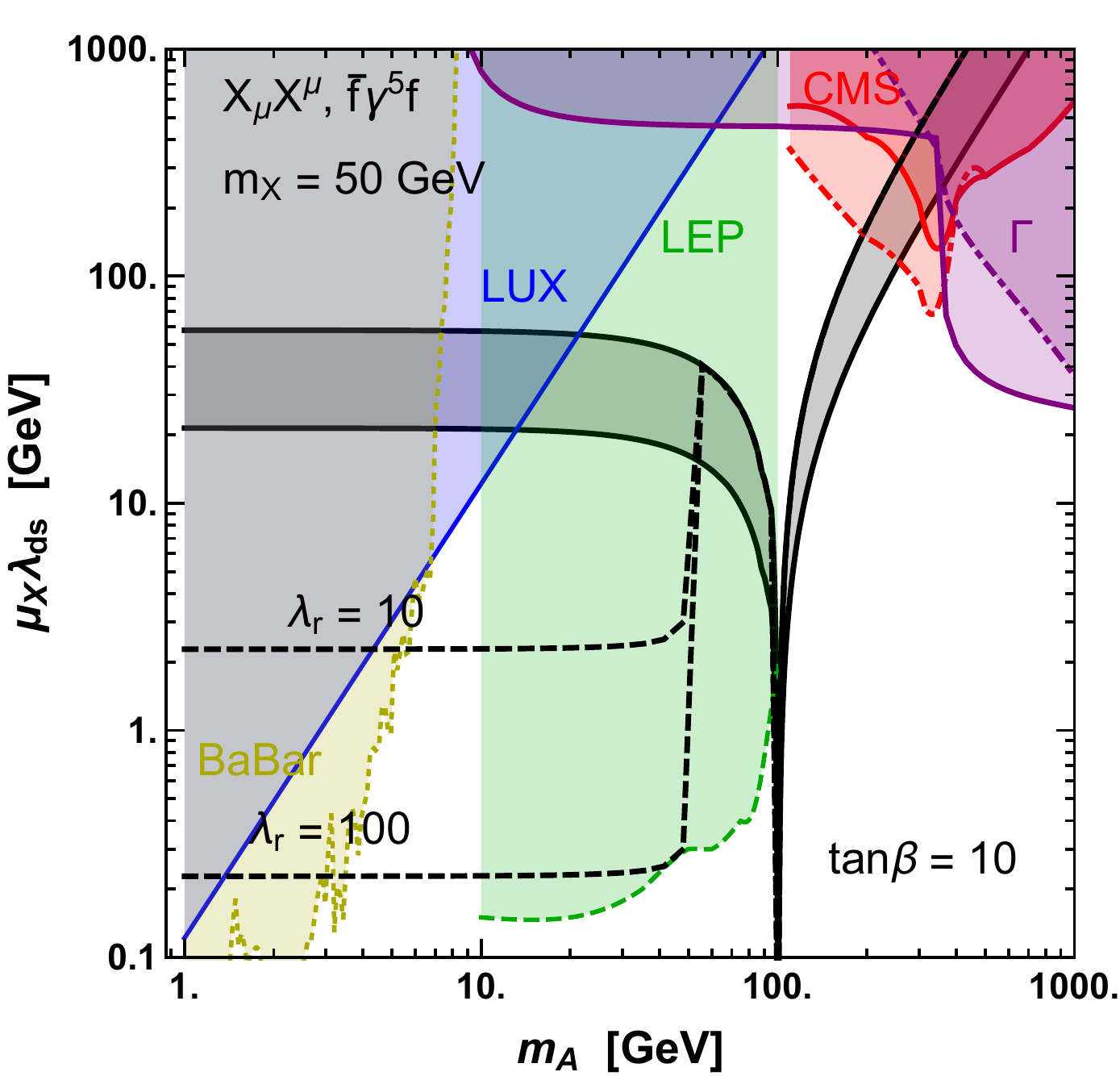}

\caption{\label{fig:spin0med_vectordm} As in previous figures, but for a 50 GeV complex (left) and real (right) vector dark matter candidate which annihilates through a spin-$0$ mediator with a pseudoscalar coupling to SM fermions. In the upper frames, we take the mediator's couplings to be equal for all SM fermions, whereas in the lower frames the mediator does not couple to leptons and $\tan \beta = 10$.}
\end{figure*}

Next, we consider the case of vector dark matter $X^\mu$:
\begin{equation}
\Lag \supset \left[ a \mu_X X^\mu X_\mu^\dagger + \sum_{f} y_{f}\bar{f}  \lambda_{f p}i \gamma^{5} f \right] A \, ,
\end{equation}
where $a = 1 (1/2)$ for a complex (real) dark matter particle.

Constraints on this model are shown in~\Fig{fig:spin0med_vectordm}. The dominant decay mode of the mediator in this model, and thus the most constraining LHC search, depends on the mass of the mediator. For $m_A \simeq 100$ GeV the dominant decay is to dark matter, and thus the most constraining search is that based on mono-jet+MET events. This picture is very different for larger mediator masses, however, for which constraints based on searches for Higgs bosons decaying to $\tau^+\tau^-$ become more stringent. Both of these search channels significantly exclude mediator masses above $100$ GeV in this class of models, for both $\lambda_r = 3$ and $\lambda_r = 1/3$. Similar to in the scalar dark matter case, however, we can relax some of these constraints by suppressing the mediator's couplings to leptons and/or by increasing $\tan \beta$ (as shown in the lower frames of Fig.~\ref{fig:spin0med_vectordm}).



\section{Vector Mediated Dark Matter \label{sec:vectormed}}

\begin{figure*}
\center
\includegraphics[width=0.49\textwidth]{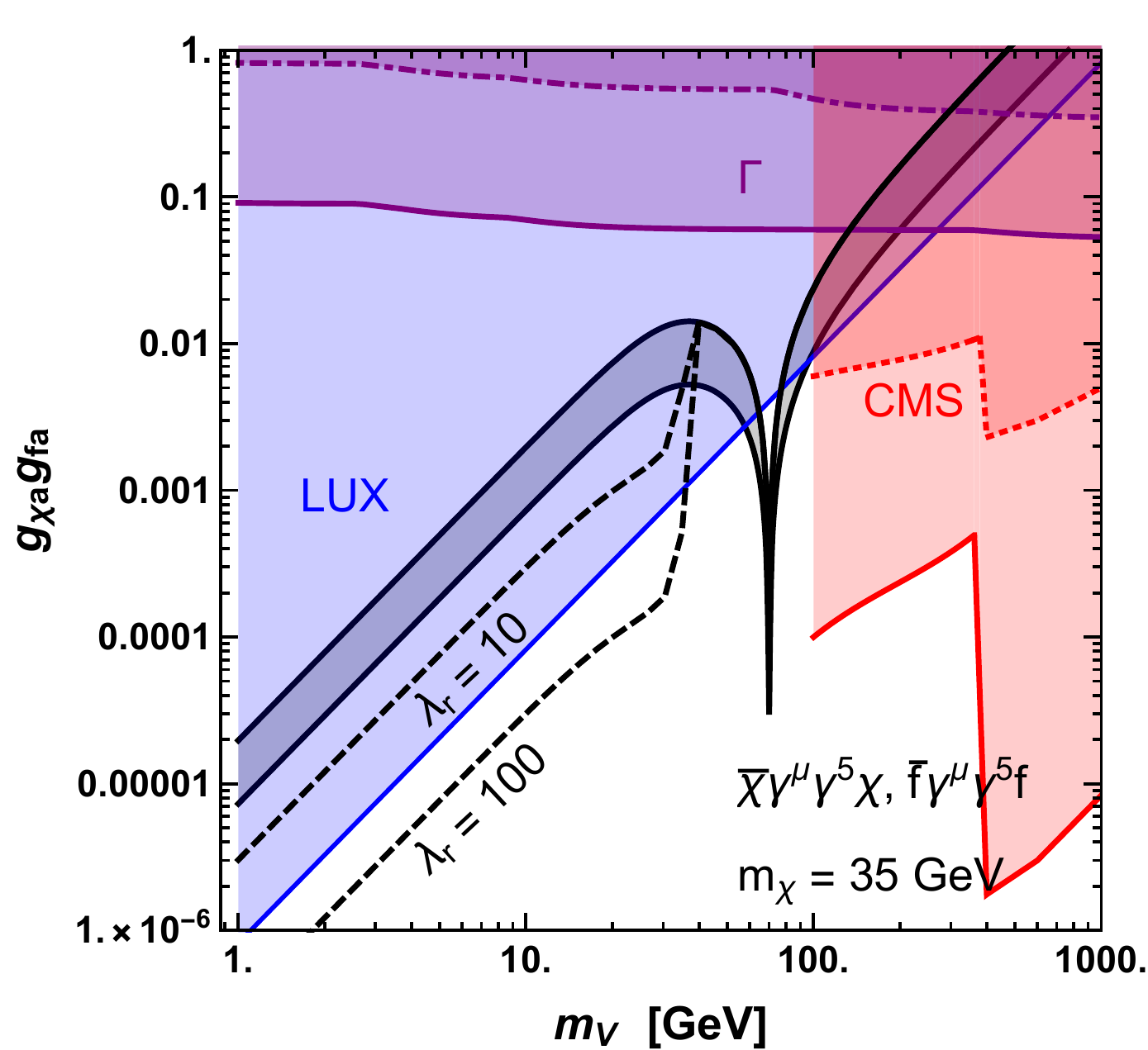}
\includegraphics[width=0.49\textwidth]{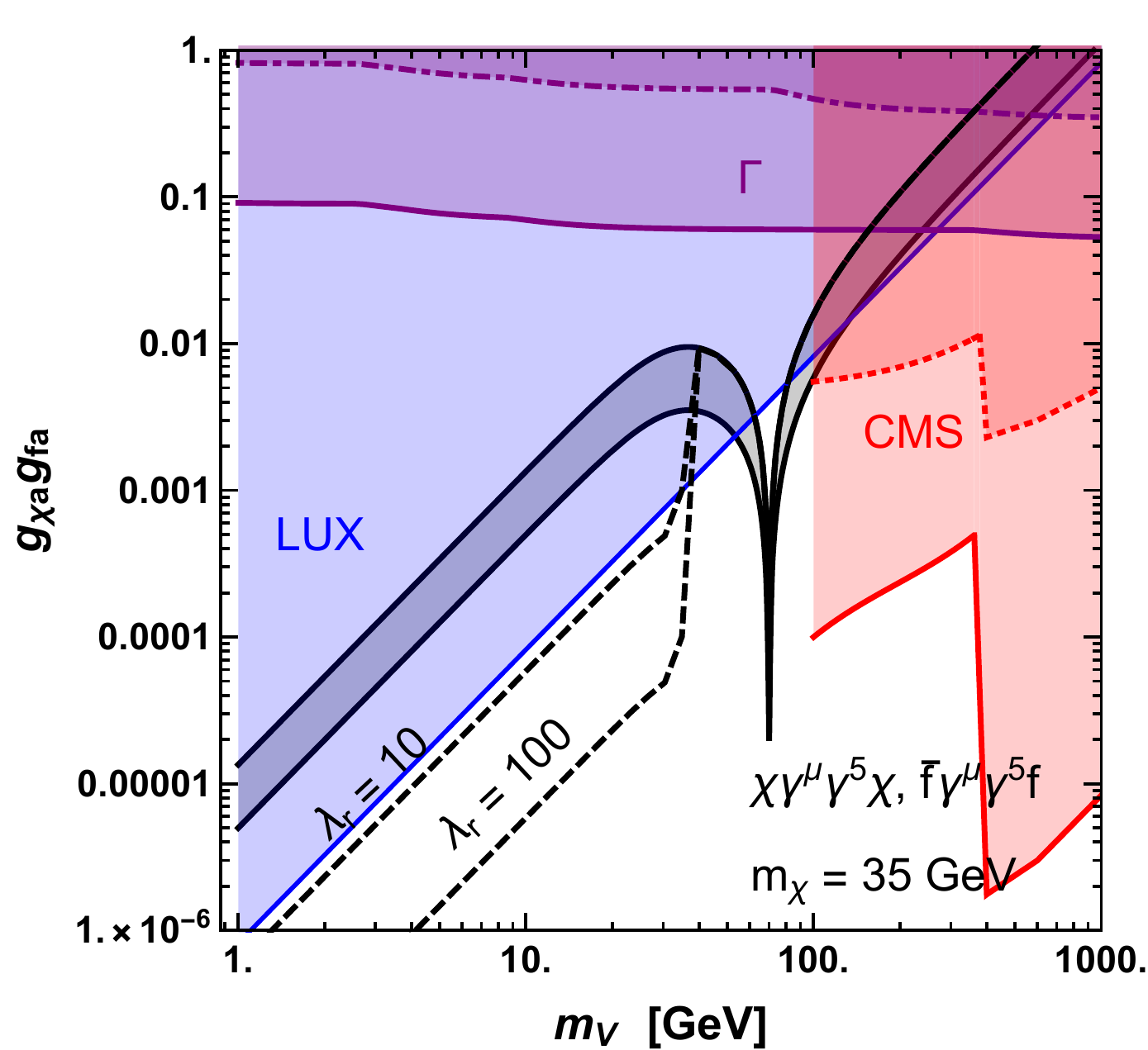}
\caption{\label{fig:spin1med_av_av} As in previous figures, but for a 35 GeV Dirac (left) and Majorana (right) dark matter candidate which annihilates through a spin-1 mediator with axial couplings to both dark matter and (universally) to SM fermions. In this figure, the dotted red (CMS) line corresponds to the case of $g_{\chi v} =1$ (\ie $\lambda_r >> 1$).}
\end{figure*}

\begin{figure*}
\center
\includegraphics[width=0.49\textwidth]{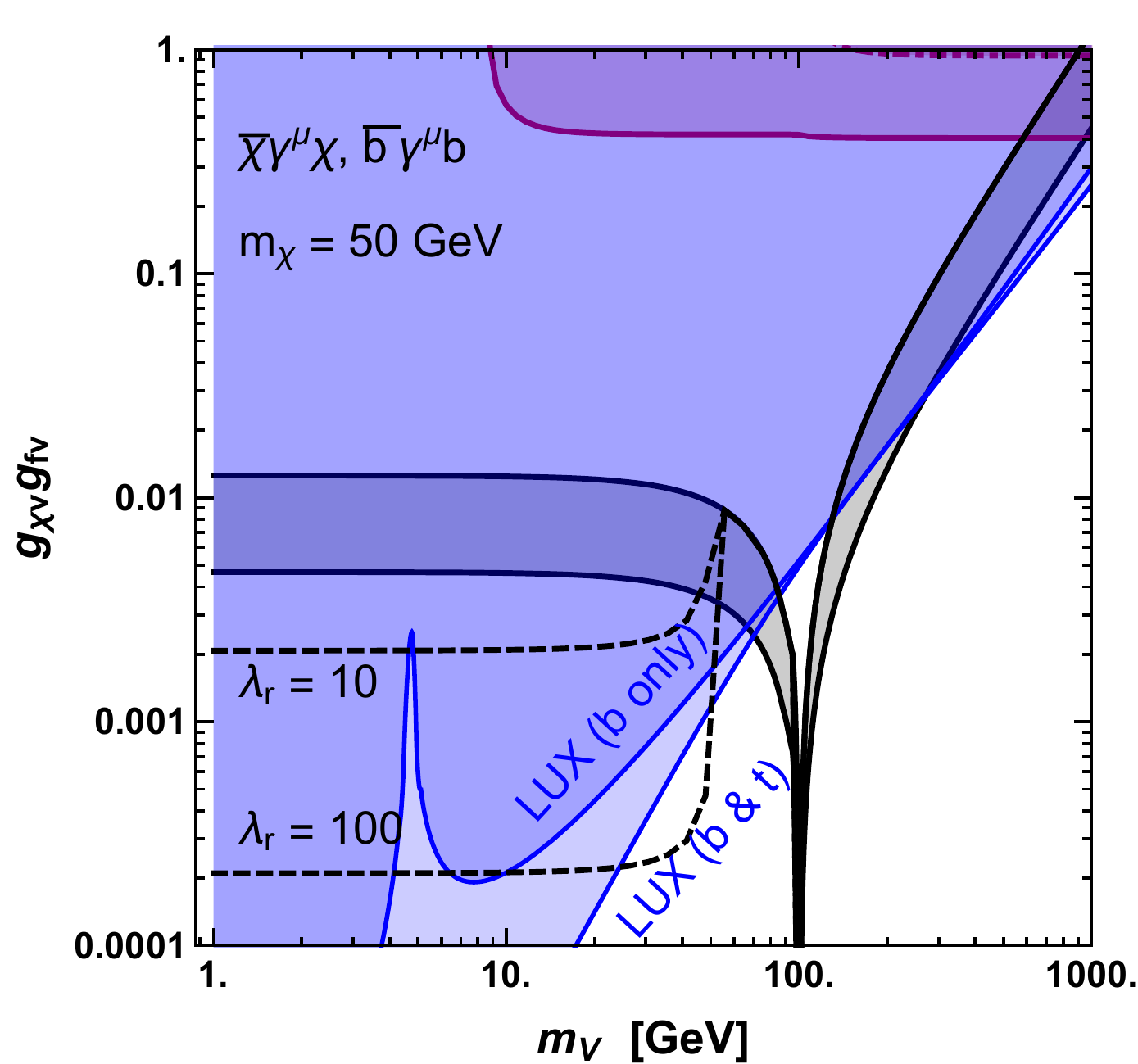}
\includegraphics[width=0.49\textwidth]{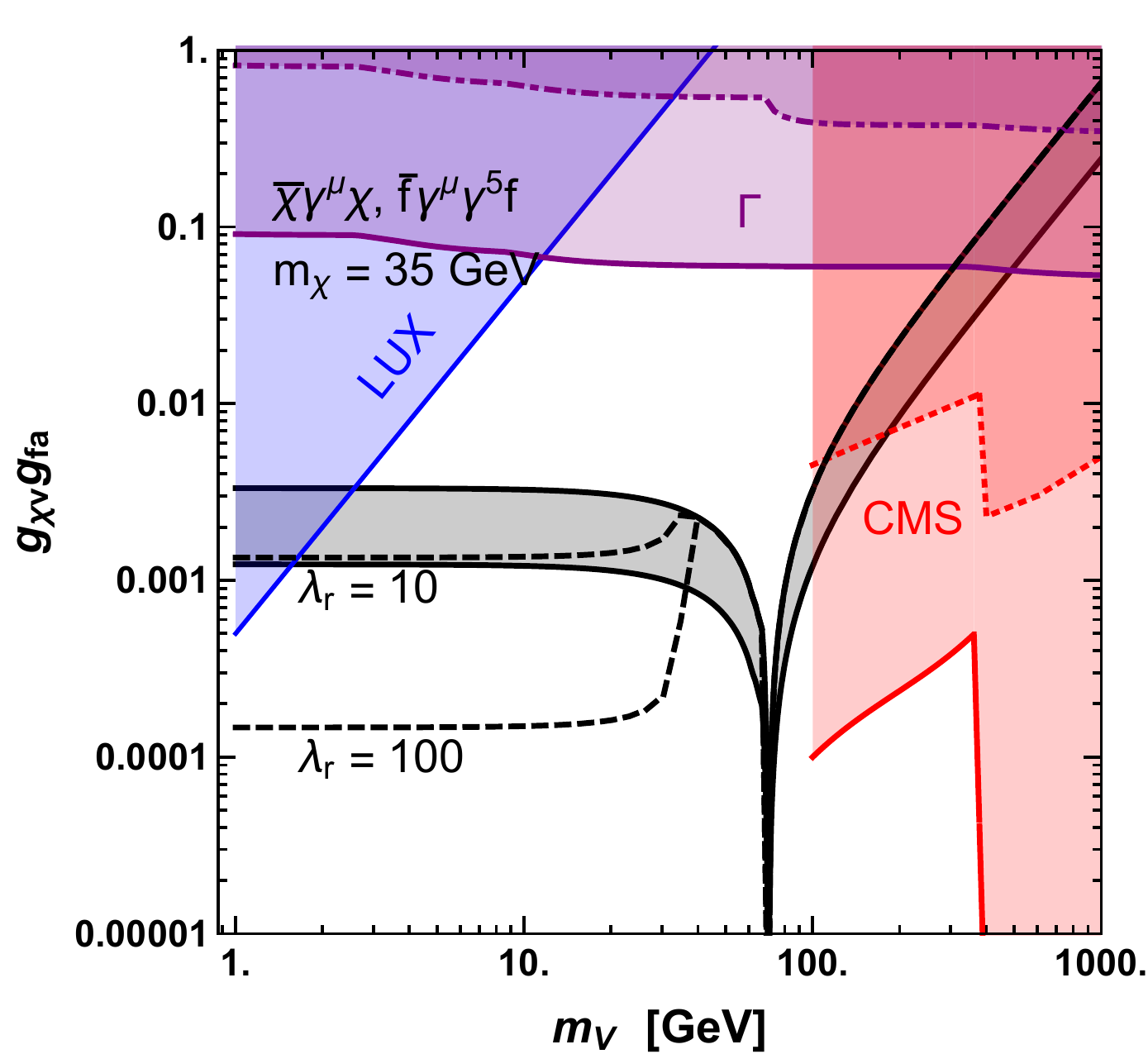}
\caption{\label{fig:spin1med_vv_vav} 
As in previous figures, but for a 50 GeV Dirac dark matter candidate with vector couplings to both dark matter and to $b$-quarks (left), and for a 35 GeV Dirac dark matter candidate with vector and axial couplings to dark matter and (universally) to SM fermions, respectively (right). LHC bounds are shown for $\lambda_r = 1/3$ (solid) and $g_{\chi v} = 1$ (\ie $\lambda_r \gg 1$) (dotted).}
\end{figure*}

In this section we consider fermonic dark matter annihilating through the $s$-channel exchange of a spin-$1$ mediator, $V_\mu$, with  
%
Lagrangians of the form~\cite{Berlin:2014tja,Hooper:2014fda}: 
\begin{equation}
\Lag \supset \left[ a\bar{\chi}\gamma^\mu(g_{\chi v} + g_{\chi a} \gamma^{5})\chi + \sum_{f}\bar{f}\gamma^\mu(g_{f v} + g_{f a} \gamma^{5})f \right] V_\mu \, ,
\end{equation}
where $a = 1 (1/2)$ for a Dirac (Majorana) dark matter candidate. For the case of comparable couplings to various SM fermions this class of models require a $\simeq 35$ GeV dark matter candidate to generate a signal consistent with the Galactic Center excess. Unless stated otherwise, we will adopt this value for the dark matter mass throughout this section. 

We begin in Fig.~\ref{fig:spin1med_av_av} by considering the constraints on a Dirac (left) and Majorana (right) dark matter candidate that annihilates through a mediator with purely axial couplings. As spin-dependent elastic scattering with nuclei is unsuppressed in this class of models\footnote{At the nuclear scale, one would in general expect vector couplings to arise at the one-loop level, leading to stronger direct detection bounds~\cite{DEramo:2016gos}. It may be possible to avoid this in some models, however, and thus we conservatively plot direct detection bounds assuming only axial couplings.}, current LUX constraints force such models to either live on resonance ($m_{\chi} \simeq m_V/2$), or have a mediator mass $m_V < m_\chi$ and with $\lambda_r \gg 1$. LHC constraints on this model from searches for di-lepton resonances ($m_V > 400$ GeV) and mono-jet+MET searches ($100 \, {\rm GeV} < m_V < 400$ GeV) limit mediator masses in this model to be below $\simeq 100$ GeV. LHC bounds are shown in this figure for $\lambda_r = 1/3$ (solid) and $g_{\chi v} = 1$ (\ie $\lambda_r \gg 1$) (dotted). Collider constraints for this model are difficult to evade as they do not rely exclusively on couplings to leptons or to specific species of quarks. Such bounds could be evaded, however, if the mediator were to couple exclusively to third generation quarks. An example of such a model is shown in the left frame of \Fig{fig:spin1med_vv_vav}, where we consider a $50$ GeV Dirac dark matter candidate that annihilates through a spin-1 mediator with vector couplings to both dark matter and $b$-quarks (and possibly also $t$-quarks). While the leading order elastic scattering diagram arises at loop level in this case, the vector coupling leads to stringent constraints from direct detection experiments. The dominant constraints from the LHC on vector mediated models typically arise from searches for mono-jet+MET events and di-lepton resonances. Since the production of the vector mediator is in most cases dominated by valence quarks, however, the sensitivity of collider searches is heavily suppressed and thus do not probe significant parameter space in this model. We do not show any LHC constraints in this figure.

In the right panel of \Fig{fig:spin1med_vv_vav}, we consider the phenomenology of models where the mediator couples to Dirac dark matter and fermions with a vector and an axial coupling, respectively. The elastic scattering cross section in this case is both spin-dependent and momentum suppressed, and thus such experiments have only recently begun probing this model. LHC constraints from di-lepton resonances ($m_V > 400$ GeV) and mono-jet+MET searches ($100 \, \rm{GeV} < m_V < 400$ GeV) are, as before, extremely constraining. That being said, di-lepton constraints can be easily avoided if the mediator couples only to quarks, and mono-jet constraints can be significantly relaxed if the mediator couples, for example, only to the third generation. LHC bounds are shown for $\lambda_r = 1/3$ (solid) and $g_{\chi v} = 1$ (\ie $\lambda_r \gg 1$) (dotted).

\section{Dark Matter Annihilating Through $t$-Channel Mediators}

Finally, we consider four scenarios in which the dark matter annihilates through the $t$-channel exchange of a colored and electrically charged mediator to $b\bar{b}$~\cite{Batell:2013zwa,Agrawal:2011ze,Agrawal:2014una}. These cases consist of a Dirac dark matter candidate, $\chi$, and spin-$0$ mediator, $A$:
\begin{equation}
\Lag \supset \lambda_{\chi} \bar{\chi}(1 + \gamma^{5})f A + \lambda_{\chi} \bar{f}(1- \gamma^{5})\chi A^\dagger \, ,
\end{equation}
a Dirac dark matter candidate, $\chi$, and a spin-$1$ mediator, $V_\mu$:
\begin{equation}\label{eq:tchan_Dirac_vec}
\Lag \supset g_{\chi} \bar{\chi}\gamma^\mu(1 + \gamma^{5})f V_\mu + g_{\chi} \bar{f} \gamma^\mu (1 - \gamma^{5})\chi V_\mu^\dagger \,
\end{equation}
and a real or complex vector dark matter candidate, $X_\mu$, with a fermionic mediator, $\psi$:
\begin{equation}\label{eq:tchan_vec_dm}
\Lag \supset g_X\bar{\psi}\gamma^\mu(1 + \gamma^{5})f X_\mu^\dagger + g_X \bar{f}\gamma^\mu(1 - \gamma^{5})\psi X_\mu \, .
\end{equation}

Note that we consider these specific combinations of scalar and pseudoscalar or vector and axial couplings as they are the only examples for which the scalar contact interaction with nuclei is supressed. Instead, elastic scattering occurs in each of these models through a loop-suppressed vector coupling~\cite{Berlin:2014tja,Agrawal:2014una,Agrawal:2010fh}.

\begin{figure*}
\center
\includegraphics[width=0.49\textwidth]{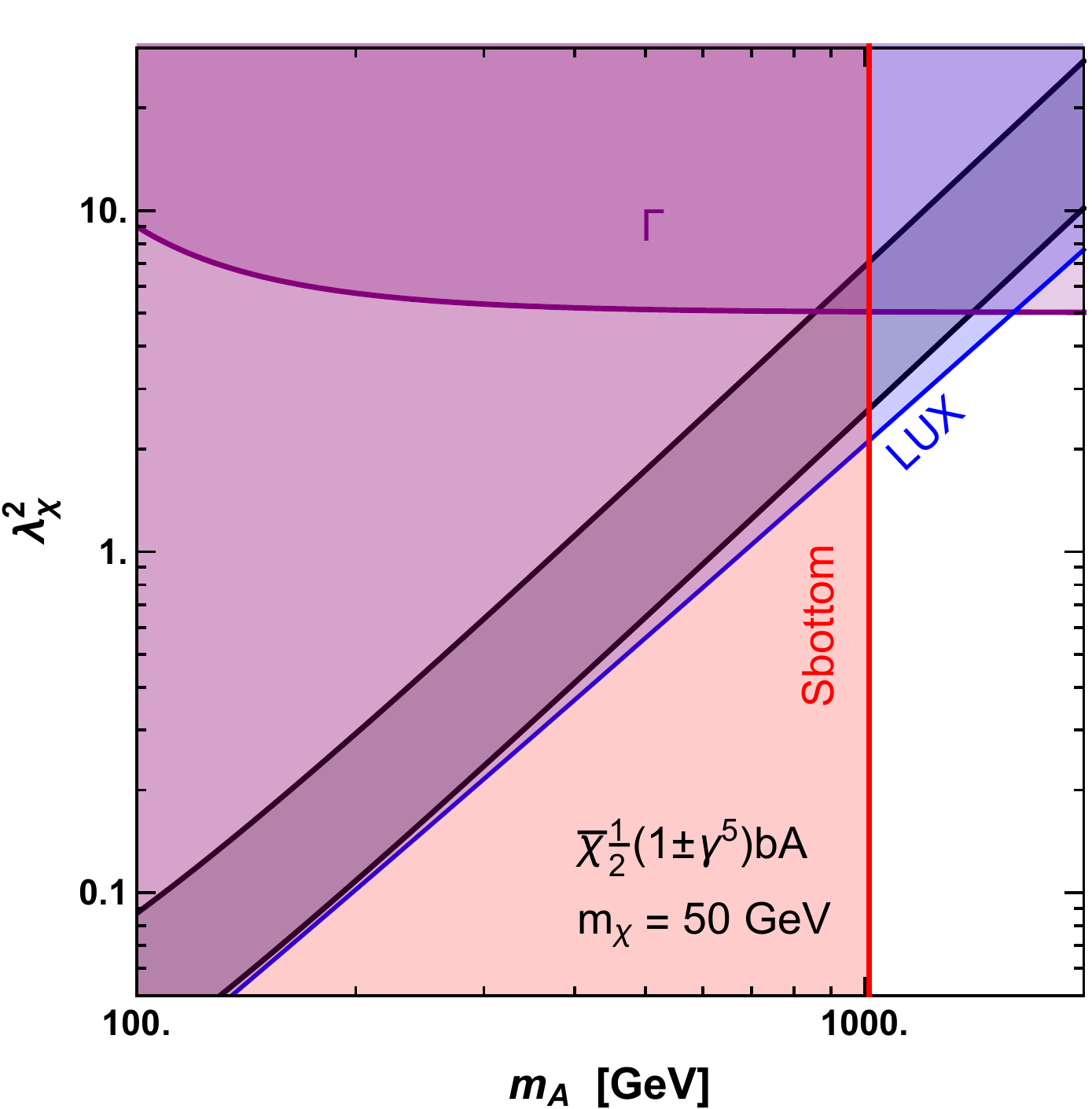}
\includegraphics[width=0.49\textwidth]{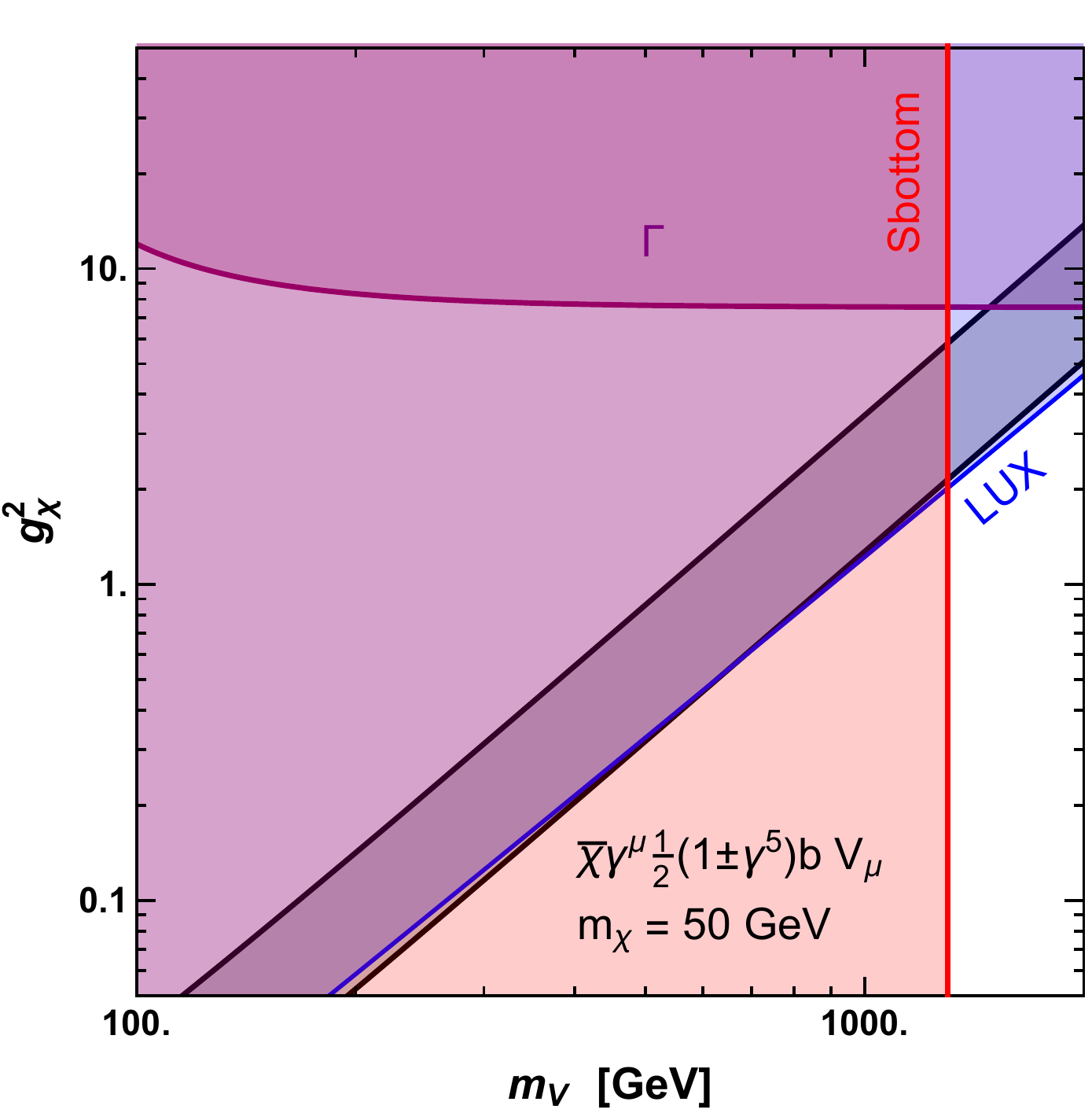}\\
\includegraphics[width=0.49\textwidth]{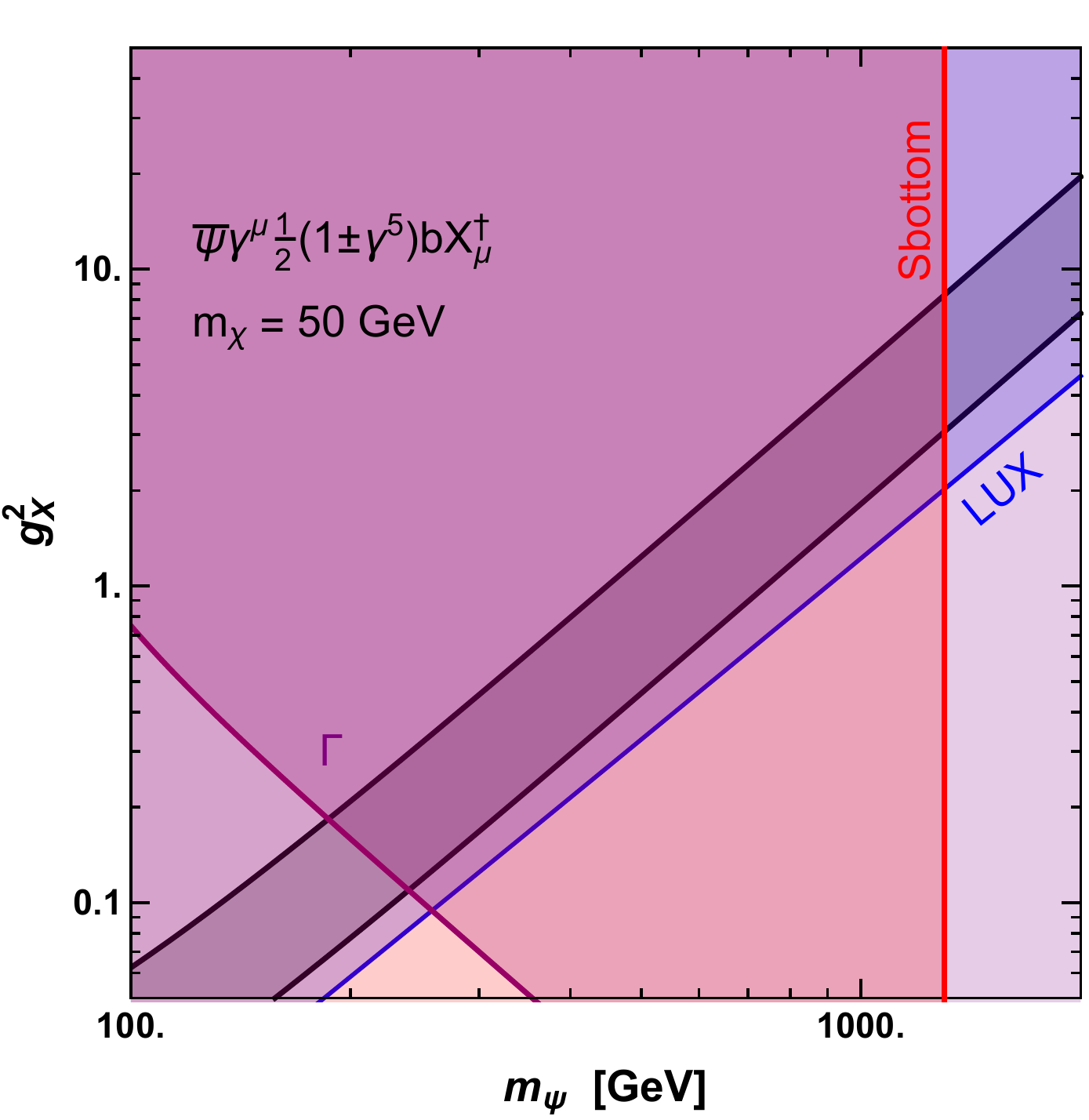}
\includegraphics[width=0.49\textwidth]{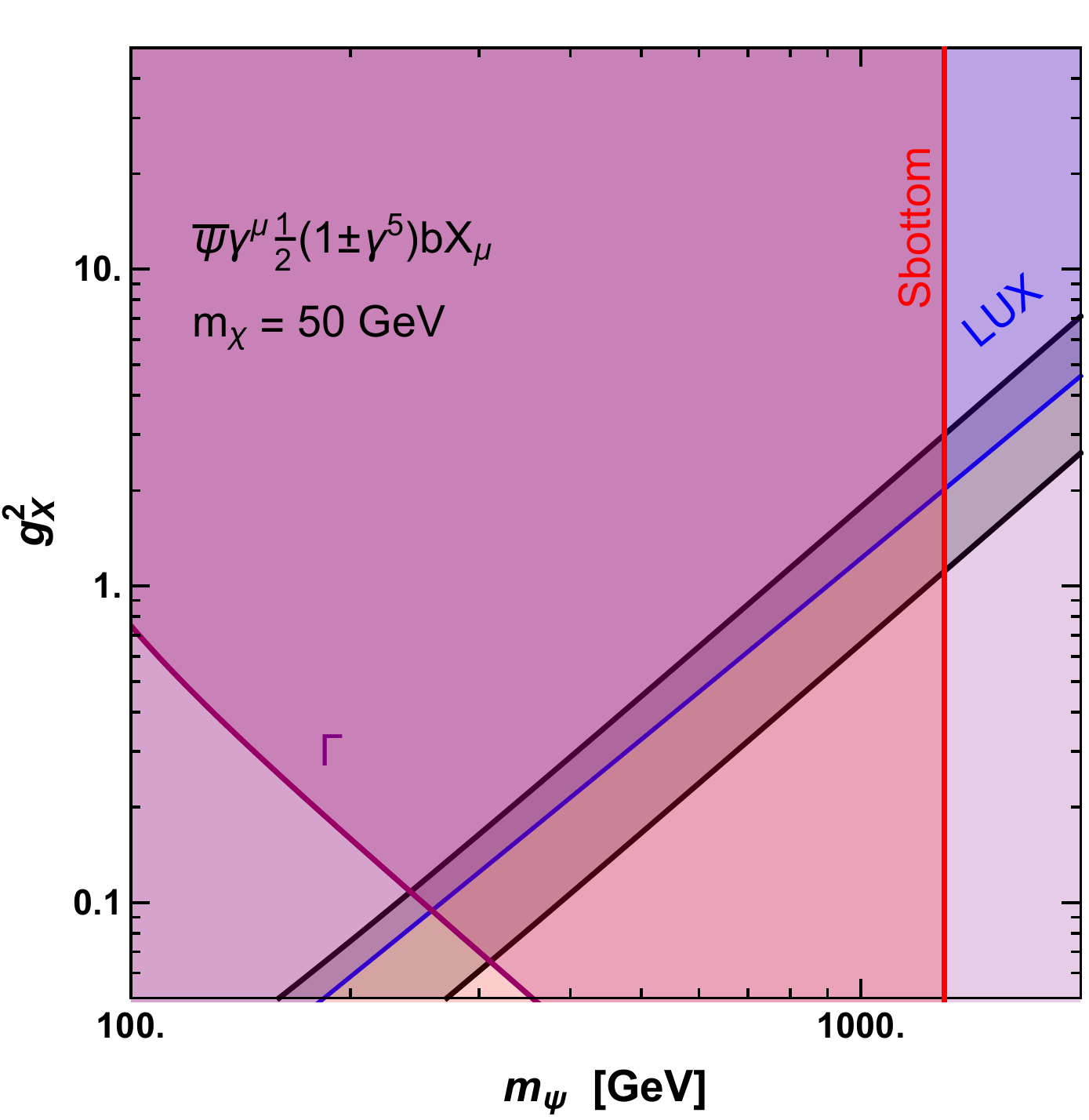}
\caption{\label{fig:tchannel_sbottom} As in previous figures, but for a 50 GeV dark matter candidate which annihilates through a $t$-channel diagram to $b\bar{b}$. In the upper left (right) frame, we consider the case of a Dirac dark matter candidate with a scalar (vector) mediator. In the lower left (right) frame, the dark matter is a real (complex) vector, mediated by a Dirac fermion. The entire parameter space of these models is ruled out by the combined results of LUX and sbottom searches at the LHC.}
\end{figure*}

In Fig.~\ref{fig:tchannel_sbottom}, we summarize the phenomenology of this class of models. In the upper left frame we consider the case of a Dirac dark matter particle and spin-0 mediator. 
In the remaining frames of this figure, we summarize the phenomenology of models with a Dirac dark matter candidate and a vector mediator (upper right), a complex vector dark matter candidate with a fermonic mediator (lower left), or a a real vector dark matter candidate with a fermonic mediator (lower right). In each case, we find that the combination of constraints from the CMS sbottom search and LUX exclude the entire parameter space of this class of models. We also note that the scenarios with a vector dark matter candidate are rather unphysical over much of the parameter space shown due to the very large  width of the mediator. 

\newpage

\section{Summary and Conclusions}

In this study, we have revisited the range of dark matter scenarios that could potentially generate the observed characteristics of the Galactic Center gamma-ray excess, without conflicting with any constraints from colliders or direct detection experiments. We have taken a simplified models approach, considering the 16 scenarios that were previously found to be viable in Ref.~\cite{Berlin:2014tja} (and listed in Table~\ref{table1}).  Each of these models features a low-velocity dark matter annihilation cross section that is unsuppressed (\ie $s$-wave), and was found to be consistent with all constraints as of 2014. Note that we have not considered any hidden sector models (\ie models in which the dark matter annihilates into unstable particles without sizable couplings to the Standard Model) which, although potentially viable~\cite{Abdullah:2014lla,Martin:2014sxa,Hooper:2012cw}, are beyond the scope of this work.

The main results of this study can be summarized as follows:
\begin{itemize}
\item{Scalar, fermonic, or vector dark matter that annihilates through a mediator with pseudoscalar couplings can in many cases evade all current constraints, for mediator masses between $\sim$10 GeV and several hundred GeV.}
\item{Dark matter that annihilates through a spin-1 mediator is ruled out by the results of LUX/PandaX-II unless the mass of the mediator is approximately equal to twice the mass of the dark matter (near an annihilation resonance). An exception to this conclusion is found in the case of a mediator with a purely vector coupling to the dark matter and a purely axial coupling to Standard Model fermions, which is potentially viable for mediator masses between roughly $\sim1$ GeV and 200 GeV.}
\item{All scenarios in which the dark matter annihilates through a $t$-channel process are now ruled out by a combination of constraints from LUX/PandaX-II and the LHC.}
\item{Constraints from LEP-II and BaBar restrict many of the pseudoscalar mediated scenarios considered in this study. In particular, mediators with a mass in the $\sim$10-100 GeV range are often ruled out by LEP if they couple significantly to the Standard Model $Z$ (such as in scenarios in which the mediator obtains its couplings to Standard Model fermions through mixing with the Higgs).}
\end{itemize}

Dark matter scenarios that are capable of generating the Galactic Center excess are now significantly more constrained than they were even a few years ago. As the sensitivity of XENON1T, LZ and other direct detection experiments, as well as the LHC, continues to improve, either a discovery will be made, or the vast majority of the currently viable parameter space identified in this study will be excluded. If such searches do advance without the appearance of new signals, hidden sector scenarios will become increasingly attractive, in particular within the context of the Galactic Center.

\bigskip

\textbf{Acknowledgments.} We would like to thank Paddy Fox, Roni Harnik, Veronica Sanz, Nuria Rius, and Asher Berlin for helpful discussions. ME is supported by the Spanish FPU13/03111 grant of MECD and also by the European projects H2020-MSCA-RISE-2015 and H2020-MSCA-ITN-2015/674896-ELUSIVES. DH is supported by the US Department of Energy under contract DE-FG02-13ER41958. SW is supported under the University Research Association (URA) Visiting Scholars Award Program, and by a UCLA Dissertation Year Fellowship. Fermilab is operated by Fermi Research Alliance, LLC, under Contract No. DE-AC02-07CH11359 with the US Department of Energy.

\bibliographystyle{JHEP}
\bibliography{gc_model_constraints}

\end{document}